\documentclass[11pt,a4paper]{article}
\pdfoutput=1

\usepackage{jcappub}
\usepackage[utf8]{inputenc}
\usepackage{aas_macros}
\usepackage{latexsym}

\newcommand\LCDM{$\Lambda$CDM}
\newcommand\Mpc{\text{Mpc}}
\renewcommand\vec{\boldsymbol}

\newcommand\code[1]{\texttt{#1}}

\title{Weak lensing of large scale structure in the presence of screening}

\author[a]{Nicolas~Tessore,}
\author[b]{Hans~A.~Winther,}
\author[a]{R.~Benton~Metcalf,}
\author[b]{Pedro~G.~Ferreira,}
\author[c]{and Carlo~Giocoli}

\affiliation[a]{Department of Physics and Astronomy, Università di Bologna, \\
viale Berti Pichat 6/2, 40127 Bologna, Italy}
\affiliation[b]{Astrophysics, University of Oxford, \\
DWB, Keble Road, Oxford, OX1 3RH, UK}
\affiliation[c]{Aix Marseille Universit\'e, CNRS, LAM (Laboratoire  d'Astrophysique de Marseille) \\
UMR 7326, 13388 Marseille, France}

\emailAdd{nicolas.tessore@unibo.it}
\emailAdd{hans.winther@astro.ox.ac.uk}
\emailAdd{robertbenton.metcalf@unibo.it}
\emailAdd{pedro.ferreira@physics.ox.ac.uk}
\emailAdd{carlo.giocoli@lam.fr}

\abstract{%
A number of alternatives to general relativity exhibit gravitational screening in the non-linear regime of structure formation.
We describe a set of algorithms that can produce weak lensing maps of large scale structure in such theories and can be used to generate mock surveys for cosmological analysis.
By analysing a few basic statistics we indicate how these alternatives can be distinguished from general relativity with future weak lensing surveys.
}

\keywords{%
modified gravity,
weak gravitational lensing,
cosmological simulations,
power spectrum
}

\begin{document}

\maketitle
\flushbottom


\section{Introduction}

A possible approach to explaining the observational evidence for cosmic acceleration is that general relativity is modified in regions of low density, low acceleration or, simply, on large scales.
A number of models have been proposed \cite{2012PhR...513....1C}, and the idea that there may be a {\it gravitational} solution to the dark energy problem has led to a renewed scrutiny of the fundamental properties of gravity.
The hope is that, in the very least, a better understanding of general relativity will emerge from current and future cosmological observations.

A common feature in a number of modified theories of gravity is that there is some form of {\it screening} \cite{2013arXiv1312.2006K}.
Precision tests \cite{2014LRR....17....4W,2015arXiv150107274B} indicate that general relativity is at least an excellent effective theory for describing gravitational physics locally or in regions of moderately high density and curvature.
This means that, in any modification of the theory, the nature of gravity must depend on its environment and that deviations from general relativity are absent (or screened) in regions in which it has been well tested.
Current proposals involve the chameleon \cite{2004PhRvD..69d4026K,2007PhRvD..75f3501M}, symmetron \cite{2010PhRvL.104w1301H,2005PhRvD..72d3535P,2008PhRvD..77d3524O} or Vainshtein \cite{1972PhLB...39..393V} mechanism, all of which lead to particular features in the gravitational potential.

A key characteristic of the various screening mechanisms \cite{2004PhRvD..69d4026K,2010PhRvL.104w1301H,1972PhLB...39..393V} currently being considered is that they are fundamentally non-linear.
While they might have an effect in regimes where density perturbations are still in the linear regime, their full effect comes into play in the non-linear regime of gravitational collapse.
This means that, if we are to fully understand the effect of screening in universes which closely resemble our own, we must be able to model non-linear structure formation not only accurately but also efficiently.
There has been substantial progress in numerically modelling the quasi-linear and non-linear regime of structure formation using adapted $N$-body codes \cite{2014A&A...562A..78L,2012JCAP...01..051L,2013MNRAS.436..348P,2015arXiv150606384W}, and we are now beginning to reach the level of accuracy required for what has been dubbed ``precision cosmology''.
Such accuracy is at the expense of highly intensive computer algorithms which focus on specific models and hence on a reduced subset of the full parameter space that needs to be explored.
We have recently advocated the development of approximate algorithms \cite{2015PhRvD..91l3507W} to explore a broader range of parameter space at the expense of accuracy. 

With a modified $N$-body code it is then possible to study the observational consequences of gravitational screening by generating a wide suite of simulations which in turn can be used to design future surveys.
Converting such simulations into mock surveys is a crucial step in optimising the scientific returns of cosmological experiments such as Euclid \cite{2015arXiv150104908S}, LSST \cite{2008arXiv0805.2366I}, WFIRST \cite{2013arXiv1305.5422S} and SKA \cite{2004NewAR..48..979C}.

One particular cosmological probe --- the weak lensing of galaxies by intervening gravitational potentials --- has been heralded as a particularly powerful test of gravitational physics on large scales.
It should supply complementary information to probes of the matter density field through spectroscopic and photometric galaxy redshift surveys and, in principle, will be insensitive to galaxy bias.
Current measurements of weak lensing on large scales are not yet competitive with other cosmological probes, but the future looks promising with the surveys mentioned above.

In this paper, we describe the algorithms that can be used to produce lensed maps in cosmologies with screening.
To produce modified gravity simulations we rely on the code described in \cite{2014A&A...562A..78L} and for lensing we use the \code{GLAMER} pipeline \cite{2014MNRAS.445.1942M,2014MNRAS.445.1954P}.
The methods we propose here are the starting point for developing a concerted and comprehensive search for signatures of screening in observations of weak lensing.
Our work complements the work developed in \cite{2014MNRAS.445.1942M,2014MNRAS.445.1954P,2015MNRAS.452.2757G,2002PhRvD..65f3001H,2009ApJ...701..945S,2000ApJ...535L...9C}.

The outline of this paper is as follows.
In Section~\ref{theory} we briefly summarise the theoretical framework with a particular and systematic focus on the different mechanisms for gravitational screening and how they can be parametrised.
In Section~\ref{sims} we describe the modified $N$-body algorithm with which we produce the density field and the lensing pipeline, \code{MapSim} and \code{GLAMER}.
In Section~\ref{results} we present the results of a suite of simulations, focusing on maps and a few of the main one point and two point statistics.
Finally, in Section~\ref{discussion} we discuss what we have found as well as the limitations and future applications of our approach. 


\section{Theory}
\label{theory}

We now establish a useful unified framework for describing the theories of gravity we have studied in this paper.

\subsection{Scalar-tensor theories}

In this paper we will focus on scalar-tensor theories of modified gravity that display some form of a screening mechanism.
These are encompassed by the general Lagrangian\footnote{%
$f(R)$ gravity can be brought to this form by performing a conformal transformation $g_{\mu\nu}\to g_{\mu\nu}A^2(\phi)$ with $A(\phi) = e^{\frac{\beta\phi}{M_{\rm Pl}}}$ and $\beta =1/\sqrt{6}$, see e.g.\@ \cite{2008PhRvD..78j4021B}.
}
\begin{equation}\label{lagr}
\mathcal{L} = \frac{R}{2}M_{\rm Pl}^2 + \mathcal{L}_\phi(\phi,\partial\phi,\partial\partial\phi) + \mathcal{L}_m(A^2(\phi)g_{\mu\nu},\psi_m),
\end{equation}
where $M_{\rm Pl}^2 \equiv (8\pi G)^{-1}$ and $\psi_m$ are the matter-fields.
Matter is coupled to the scalar field via the effective metric
\begin{equation}
\tilde{g}_{\mu\nu} = A^{2}(\phi)g_{\mu\nu},
\end{equation}
and therefore moves on geodesics of the $\tilde{g}_{\mu\nu}$ metric.
In the Einstein frame, where gravity is described by general relativity, this implies the existence of a fifth-force; for a test mass in the non-relativistic limit the force is given by
\begin{equation}\label{fifth_force_eq}
{\vec F}_\phi = {\vec \nabla}\log A(\phi).
\end{equation}
The key to a successful modified gravity model is to have a way of evading the stringent constraints from local gravity experiments and at the same time give rise to interesting astrophysical and cosmological signatures.
One such way is through what is commonly referred to as a screening mechanism and below we will give a short review of the different screening mechanisms we investigate in this paper.


\subsection{Screening mechanisms}

To see how screening can emerge in a scalar tensor theory let us expand the general scalar field Lagrangian Eq.~\eqref{lagr} to quadratic order about a given field value $\phi_0$
\begin{equation}\label{eq:lagrange}
\mathcal{L}_{\rm quadratic} = X^{\mu\nu}(\phi_0)\partial_{\mu}\delta\phi\partial_{\nu}\delta\phi + m^2(\phi_0)\delta\phi^2 + \frac{\beta(\phi_0)\delta\phi}{M_{\rm Pl}}\rho_m,
\end{equation}
where $m(\phi_0) = \sqrt{d^2V(\phi_0)/d\phi^2}$ is the (local) value of the scalar field mass, $V(\phi)$ is the potential for $\phi$, and $\beta = \frac{d\log A}{d\phi}M_{\rm Pl}$ is the (local) coupling strength of the fifth-force.
The value $2\beta^2=1$ corresponds to a force with the strength of gravity.
The fifth-force on a test mass, from a source of mass $M$, can schematically be written in the form
\begin{equation}
F_\phi \simeq \frac{GM}{r^2} \frac{2\beta^2(\phi_0)}{\sqrt{|X^{\mu\nu}(\phi_0)|}} e^{-m(\phi_0) r}.
\end{equation}

Let us now consider two different regions of space-time, one where $\phi_0 = \phi_{A}$ and another where $\phi_0 = \phi_{B} \not= \phi_A$ respectively.
From the quadratic Lagrangian~\eqref{eq:lagrange} we can see that there are at least three ways that we can have some form of screening.
One way to reduce the effect of the fifth-force (in region B compared to A) is by having a large local mass $m(\phi_B) \gg m(\phi_A)$ which implies a very short interaction range (the chameleon mechanism).
If the matter coupling $\beta(\phi_B) \ll \beta(\phi_A)$, the fifth-force will also be suppressed (the symmetron mechanism).
Lastly, $|X^{\mu\nu}(\phi_B)| \gg |X^{\mu\nu}(\phi_A)|$ leads to, after canonical normalisation, a weakened matter source and therefore also a weakened fifth-force (the Vainshtein mechanism).

We should note that this simplified description does not tell the whole story; added to these effects we can (and do) have additional screening which can only be studied by considering the full non-linear dynamics of the scalar field.

 
\paragraph{Chameleon mechanism}

The chameleon mechanism \cite{2004PhRvD..69d4026K,2007PhRvD..75f3501M} can be found in models defined by the Lagrangian
\begin{equation}
\mathcal{L}_\phi = \frac{1}{2}(\nabla\phi)^2 + V(\phi). 
\end{equation}
The Klein-Gordon equation for the scalar field becomes 
\begin{equation}
\Box\phi + V_{\text{eff},\phi} = 0.
\end{equation}
In the presence of matter sources, the dynamics of $\phi$ are determined by an effective potential which, for non-relativistic matter, is given by
\begin{equation}
V_{\text{eff}} =  V(\phi) +  \frac{A(\phi)\rho_m}{M_{\rm Pl}}~.
\end{equation}
Too see how screening works in detail, we look at a static, spherically symmetric object of density $\rho_0$ and radius $R$ embedded in a background of density $\rho_{\rm env}$.
The approximate solution to the Klein-Gordon equation \cite{2004PhRvD..69d4026K} in this case leads to a fifth-force given by
\begin{equation}\label{force_cham}
F_\phi \simeq 2\beta^2 \frac{\Delta R}{R} \frac{GM}{r^2} e^{-m_{\rm env} r},
\end{equation}
where $m_{\rm env} = \sqrt{V_{\rm eff,\phi\phi}(\phi_{\rm env})}$ is the mass of the scalar field in the background, $\beta = M_{\rm Pl}\frac{d\log A}{d\phi}$ is the coupling strength (which is constant in the chameleon model) and
\begin{equation}\label{screen_cond}
\frac{\Delta R}{R} = \text{min}\left\{1, \frac{|\phi_0-\phi_{\rm env}|}{2\beta M_{\rm Pl}\Phi_N}\right\}
\end{equation}
is the so-called thin-shell screening factor where $\Phi_N$ is the Newtonian potential of the object and subscript "0" ("${\rm env}$") refer to quantities at the centre (in the environment) of the object.
The larger the Newtonian potential becomes, the smaller $\frac{\Delta R}{R}$ is, and the fifth-force is screened. On top of this, and in dense environments, the term $|\phi_0-\phi_{\rm env}|$ also becomes small (and $m_{\rm env}$ becomes large), giving rise to an environmental screening effect, i.e.\@ an object that is not screened on its own can be screened if it is in a very dense environment.
This environmental dependence has many interesting consequences, see e.g.\@ \cite{2010PhRvD..81j3002S,2012ApJ...756..166W,2011PhRvL.107g1303Z}.


\paragraph{Symmetron mechanism}

The original symmetron model \cite{2010PhRvL.104w1301H} is defined by the same Lagrangian as for the chameleon where 
\begin{gather}
V(\phi) = -\frac{1}{2}\mu^2\phi^2 + \frac{1}{4}\lambda\phi^4,\\
A(\phi) = 1 + \frac{1}{2}\frac{\phi^2}{M^2} \to \beta(\phi) = \frac{\phi M_{\rm Pl}}{M^2},
\end{gather}
and where $\mu,M,\lambda$ are model parameters. The effective potential in the presence of matter sources is given by
\begin{equation}
V_\text{eff}(\phi) = \frac{1}{2}\left(\frac{\rho_m}{\mu^2M^2} -1\right)\mu^2 \phi^2 + \frac{1}{4}\lambda\phi^4.
\end{equation}
The symmetron mechanism is similar to the chameleon mechanism, except that the coupling $\beta$ is now field dependent, leading to an additional screening effect.
If the local density satisfies $\rho_m > \mu^2 M^2$, the effective potential has a minimum at $\phi = 0$ near which the field will reside.
Since the coupling is proportional to $\phi$, the effective matter coupling is suppressed in high density regions and the fifth-force is additionally screened. 

The same screening condition Eq.~\eqref{force_cham}--\eqref{screen_cond} as we had for the chameleon also applies for the symmetron, but now $\beta = \beta(\phi_{\rm env})$ is not a constant anymore. 


\paragraph{Vainshtein mechanism}

The Vainshtein mechanism \cite{1972PhLB...39..393V} is responsible for the viability of \emph{massive gravity}, but it can be present in other theories, most notably the Galileons \cite{2009PhRvD..79f4036N} and the DGP model  \cite{2000PhLB..485..208D}.
We will not work directly with the original DGP model, but instead take a toy model that has the same fifth-force, with the corresponding Vainshtein screening, but where the background is that of \LCDM.
This is known as the normal branch DGP model where we have added dark energy in the form of a cosmological constant \citep{2014JCAP...07..058F}.
The field equation for the scalar field, which in the DGP model is the so-called brane-bending mode and describes the curvature of the 4D brane we are confined to live in, reads 
\begin{equation}
\Box\phi + \frac{r_c^2}{3\beta_{\rm DGP}(a) a^2} ((\Box\phi)^2 - (\nabla_\mu\nabla_\nu\phi)^2)
= \frac{8\pi G a^2}{3\beta_{\rm DGP}}\delta\rho_m,
\end{equation}
where $r_c$ is the so-called cross-over scale and $a$ is the scale parameter of the background metric.
In the original DGP model, $r_c$ dictates at what length-scales gravity becomes 5D and $\beta_{\rm DGP}(a) = 1 + 2r_c H(a) \left(1 + \frac{\dot{H}}{3H^2}\right)$. For static spherical symmetric configurations, the field equation reduces to
\begin{equation}
\frac{1}{r^2}\frac{d}{dr}\left[r^2\frac{d\phi}{dr}\right] + \frac{2r_c^2}{3\beta_{\rm DGP}(a)}\frac{d}{dr}\left[r\left(\frac{d\phi}{dr}\right)^2\right] 
= \frac{8\pi G \delta\rho_m}{3\beta_{\rm DGP}(a)}.
\end{equation}
This equation can be integrated to yield 
\begin{equation}
\frac{d\phi}{r dr} + \frac{2r_c^2}{3\beta_{\rm DGP}(a) a^2}\left(\frac{d\phi}{r dr}\right)^2 = \frac{2a^2}{3\beta_{\rm DGP}(a)} \frac{GM}{r^3},
\end{equation}
which results in the fifth-force on a test-mass being
\begin{equation}
F_{\phi} = \frac{GM}{r^2} \ \frac{1}{3\beta_{\rm DGP}(a)} \ 2\left[\frac{\sqrt{1+(r_V/r)^3}-1}{(r_V/r)^3}\right],
\end{equation}
where $r_V = \left(\frac{16GM}{9\beta^2_{\rm DGP}} r_c^2 \right)^{1/3}$ is the so-called Vainshtein radius.
The screening mechanism works so that the fifth-force is screened at distances smaller than the Vainshtein radius ($r \ll r_V$).
As opposed to the chameleon and symmetron screening mechanisms, where we have a mass-term which gives the fifth-force at finite range, gravity in this case is modified on the largest scales: $F_\phi \simeq \frac{GM}{r^2}\frac{1}{3\beta_{\rm DGP}(a)}$ for $r\gg r_V$.


\subsection{Evolution of linear perturbations in modified gravity}\label{sect:linear}

The evolution of linear matter density perturbations in the models we are interested in can be written in the general form \cite{2012PhRvD..86d4015B}
\begin{equation}\label{eq:lineq}
\ddot{\delta}_m + 2H\dot{\delta}_m = \frac{3}{2}\Omega_m(a) H^2\delta_m \frac{G_\text{eff}(k,a)}{G},
\end{equation}
where $G_{\rm eff}(k,a)$ is the effective gravitational constant in Fourier space.
If $G_{\rm eff}(k,a) = G$ we recover the equation governing the evolution of the density perturbations in \LCDM.

For the chameleon and symmetron mechanism we have \cite{2012PhRvD..86d4015B}
\begin{equation}
\frac{G_\text{eff}(k,a)}{G} = 1 + \frac{2\beta^2(a)k^2}{k^2 + a^2m(a)^2},
\end{equation}
where $\beta(a)$ and $m(a)$ are the respective coupling and mass of the scalar field along the cosmological attractor solutions.
The form of this equation can be understood by noting that it is the ratio of the Fourier transform of the fifth-force potential to the Fourier transform of the Newtonian gravitational potential $\mathcal{F}[\nabla^2\Phi_N + \nabla^2\log A]/\mathcal{F}[\nabla^2\Phi_N] = 1 + 2\beta^2(a)\frac{k^2}{k^2 + m^2a^2}$.

For large scales $k/a \ll \frac{1}{m(a)}$ we have $\frac{G_{\rm eff}(k,a)}{G} \approx 1$ and we recover the \LCDM\ evolution.
On small scales, $k/a \gg 1/m(a)$, $\frac{G_{\rm eff}(k,a)}{G} = 1 +  2\beta^2(a)$ and gravity is modified.
For the symmetron we have
\begin{gather}
m(a) = \frac{1}{\lambda_\phi}\sqrt{1-(a_{\rm ssb}/a)^3},\\
\beta(a) = \beta_0\sqrt{1-(a_{\rm ssb}/a)^3},
\end{gather}
where $\lambda_\phi = \frac{1}{\sqrt{2}\mu}$ is the range of the symmetron field at $z=0$, $\beta_0 = \frac{\mu M_{\rm Pl}}{\sqrt{\lambda}M^2}$ is the coupling strength relative to gravity and $a_{\rm ssb} = \left(\frac{3H_0^2M_{\rm Pl}^2\Omega_m}{M^2\mu^2}\right)^{1/3}$ is the scale-factor at which the modifications of gravity become noticeable.

For the Hu-Sawicky $f(R)$ model \cite{2007PhRvD..76f4004H}, which is one of the models we have simulated, we have
\begin{gather}\label{eq:mbetafofr}
m(a) = \frac{H_0\sqrt{\Omega_m + 4\Omega_\Lambda}}{\sqrt{(n+1)|f_{R0}|}}\left(\frac{\Omega_m a^{-3} + 4\Omega_\Lambda}{\Omega_m + 4\Omega_\Lambda}\right)^{n/2+1},\\
\beta(a) = \frac{1}{\sqrt{6}},
\end{gather} 
where $n$ and $|f_{R0}|$ are model parameters. In this paper we will only consider the case where the primordial power spectrum index is $n=1$. The DGP model we work with corresponds to taking
\begin{gather}
m(a) = 0, \\
\beta(a) = \frac{1}{\sqrt{6\beta_{\rm DGP}(a)}}.
\end{gather}
Note that, since $m(a) = 0$, there is no scale (or $k$)  dependence in $G_{\rm eff}(a)$. 

Another way to look at these models is within the $\gamma,\mu$ parameterisation (see e.g. \cite{2014PhRvD..89b4026B}), where the metric potentials are given by $\nabla^2\Psi = 4\pi G a^2 \mu \delta\rho_m$ and $\Phi = \gamma \Psi$.
We have
\begin{gather}
\mu = \frac{m^2(a)a^2 + (1+2\beta^2(a))k^2}{m^2(a) a^2 + k^2},\\
\gamma = \frac{m^2(a)a^2 + (1-2\beta^2(a))k^2}{m^2(a) a^2 + (1+2\beta^2(a))k^2}.
\end{gather}
For the symmetron and $f(R)$ gravity the two functions interpolate between $\mu = \gamma = 1$ for large scales ($k/a \ll m(a)$) to $\mu = 1 + 2\beta^2(a)$ and $\gamma = 1 - \frac{4\beta^2(a)}{1+2\beta^2(a)}$ for small scales ($k/a \gg m(a)$).
On the other hand, in our DGP-like model we have $\mu = 1 + 2\beta^2(a)$ and $\gamma = 1 - \frac{4\beta^2(a)}{1+2\beta^2(a)}$ for all $k$, so gravity is modified even on the largest scales.

Linear theory is useful for obtaining a qualitative understanding of what signatures to expect, but it neglects an important part of the model's behaviour, namely the screening mechanism.
$N$-body simulations of modified gravity models, such as the ones considered in this paper, have shown that the predictions of linear perturbation theory become inaccurate as soon as the evolution of the density perturbations starts to deviate from \LCDM\ \cite{2012ApJ...748...61D, 2012JCAP...10..002B,2013JCAP...04..029B,2009PhRvD..80d3001S,2013JCAP...05..023L,2013JCAP...10..027B}.
To obtain accurate predictions for the non-linear evolution of these models, we therefore need $N$-body simulations.


\subsection{Gravitational lensing in modified gravity}

Gravitational lensing is determined by the so-called lensing potential $\Phi_+ = \frac{\Phi+\Psi}{2}$.
For the modified gravity models we consider in this paper this is equivalent to the Newtonian potential.
This can most easily be seen from the $\gamma,\mu$ parameterisation mentioned in the previous section, which gives the following prediction
\begin{equation}
\Phi_+ = \frac{(1+\gamma)\mu}{2}\Phi_N,
\end{equation}
where $\nabla^2\Phi_N = 4\pi G a^2 \delta\rho_m$ is the Newtonian potential, i.e.\@ the same equation as in \LCDM.

When $(1+\gamma)\mu = 2$, which is the case for the particular modified gravity theories we study in this paper, lensing itself is not modified and the only differences in lensing with respect to \LCDM\ are encoded in the differences in the matter distribution caused by the modifications of gravity during the process of gravitational collapse.
Modified gravity models where lensing itself is modified (see e.g. \cite{2013JCAP...10..027B}) can be studied within our numerical framework, but we leave this for future work.


\section{Simulations}
\label{sims}

The numerical results we obtain in this paper have two core elements which we now describe in turn: a modified $N$-body solver to generate the density field and a weak lensing pipeline that can convert a density field into a lensing map.

\subsection{Gravitational \texorpdfstring{$N$}{N}-body code}

The simulations in this paper have been performed using the \code{ISIS} code \cite{2014A&A...562A..78L} which is a modified gravity modification of the multi-purpose $N$-body code \code{RAMSES} \cite{2002A&A...385..337T}.
The DGP model has been implemented following the description laid out in \cite{2013JCAP...05..023L}.

Standard dark matter $N$-body simulations are evolved using only two equations.
First the gravitational potential is calculated (having first used the location of the $N$-body particles to calculate the density field) using the Poisson equation
\begin{equation}
\nabla^2\Phi_N = 4\pi G a^2 (\rho_m-\overline{\rho}_m),
\end{equation}
and the particles are then evolved using the geodesic equation
\begin{equation}
\vec{\ddot x} + 2H \vec{\dot x} = -\nabla\Phi_N.
\end{equation}
When going to our modified gravity models, the only change (when the background is close to \LCDM) is that we need to include the fifth-force.
This adds a term $\vec F_\phi$ (see Eq.~\eqref{fifth_force_eq}) to the right hand side of the geodesic equation.
The expression for the force terms and the corresponding field equations that we solve in the $N$-body code can be found in Appendix~\ref{field_equations}.
Solving these highly non-linear differential equations is the most challenging and time-consuming part of modified gravity simulations.\footnote{%
See \cite{2015arXiv150606384W} for a comparison of different codes used to simulate modified gravity.
The code we used here was found to compare very well (to per cent accuracy) with other high-resolution codes deep into the non-linear regime.
}
All the models we have simulated have been simulated before; for more information and details about the implementation of the scalar field solver and modified gravity simulations in general we refer the reader to \cite{2014A&A...562A..78L,2012JCAP...01..051L}.
The code we used was also part in a recent code comparison project for modified gravity $N$-body codes \cite{2015arXiv150606384W} and it was found to agree to the $\sim 1\%$ level deep into the non-linear regime ($k\sim 5~h/\Mpc$) with similar codes for the models simulated for this paper.

The simulations we have performed all start from the same initial conditions and are run with $N=512^3$ particles in a box of size $B_0 = 250.0~\Mpc/h$ and with a \LCDM\ cosmology given by $\Omega_m = 0.271$, $\Omega_\Lambda = 0.729$, $h = 0.703$, $n_s=0.966$ and $\sigma_8 = 0.8$.
For each simulation we choose to output the particles needed for the lensing analysis at redshifts $z = 0.000$, $0.046$, $0.111$, $0.176$, $0.244$, $0.333$, $0.422$, $0.538$, $0.660$, $0.818$, $1.000$, $1.250$, $1.500$, $1.750$, and $1.981$.

The modified gravity models we simulate are the Hu-Sawicky $f(R)$ model, the symmetron model, and the normal branch DGP model.
See Table~\ref{tab:runs} for the parameters used in the simulations.\footnote{%
With the parameter choices we have made the DGP model with $r_cH_0=1.2$ ($r_cH_0 = 5.6$) have the same value of $\sigma_8$ as the $f(R)$ model with $f_{R0} = 10^{-5}$ ($f_{R0} = 10^{-5}$).
}
The background evolution in all the modified gravity simulations is the same as in \LCDM, allowing for direct comparison of the effect of the fifth-force (and the corresponding screening mechanism).

\begin{table}
\centering
\begin{tabular}{|l|ll|}
\hline
Model & Type & Parameters \\
\hline
$F_5$          & $f(R)$    & $|f_{R0}|=10^{-5}$, $n=1$ \\	
$F_6$          & $f(R)$    & $|f_{R0}|=10^{-6}$, $n=1$ \\
Symm.~$A$      & Symmetron & $\lambda_{\phi} = 1.0~\Mpc/h$, $a_\text{SSB} = 0.50$, $\beta_0=1.0$ \\
Symm.~$B$      & Symmetron & $\lambda_{\phi} = 1.0~\Mpc/h$, $a_\text{SSB} = 0.33$, $\beta_0=1.0$ \\
DGP, $r = 1.2$ & DGP       & $r_c H_0 = 1.2$ \\
DGP, $r = 5.6$ & DGP       & $r_c H_0 = 5.6$ \\
\hline
\end{tabular}
\caption{%
The modified gravity models for which we have performed $N$-body simulations.
The background cosmology in all the simulations is a standard \LCDM\ cosmology with $\Omega_m = 0.271$, $\Omega_\Lambda = 0.729$, $h = 0.703$, $n_s=0.966$ and $\sigma_8 = 0.8$.
}
\label{tab:runs}
\end{table}


\subsection{Lensing pipeline}

Light passing through an inhomogeneous matter field is deflected by the intervening large-scale structure.
This effect, called \emph{cosmic shear}, promises to be a powerful probe of cosmology.
As long as $\Psi = \Phi$ holds for the gravitational potential, the cosmic deflection potential for a light cone out to comoving distance $\chi$ is given by (e.g.\@ \cite{2003moco.book.....D,2006glsw.conf.....M})
\begin{equation}\label{eq:psi_cone}
	\psi(\vec\theta) = \frac{2}{c^2} \int_0^\chi \! \frac{D_A(\chi - \chi')}{D_A(\chi) \, D_A(\chi')} \, \Phi(D_A(\chi') \, \vec\theta, \chi') \, d\chi' \;,
\end{equation}
where $D_A$ is the comoving angular diameter distance.
The deflection potential is sourced by an effective dimensionless surface mass density
\begin{equation}\label{eq:kappa_cone}
	\kappa(\vec\theta) = \frac{3 H_0^2 \Omega_m}{2 c^2} \int_0^\chi \! \frac{D_A(\chi') \, D_A(\chi - \chi')}{D_A(\chi)} \, \frac{\delta(D_A(\chi') \, \vec\theta, \chi')}{a(\chi')} \, d\chi' \;,
\end{equation}
where $\delta$ is the matter density contrast $\Delta\rho/\rho$.

\begin{figure}[tbp]%
\centering%
\includegraphics[width=.8\textwidth]{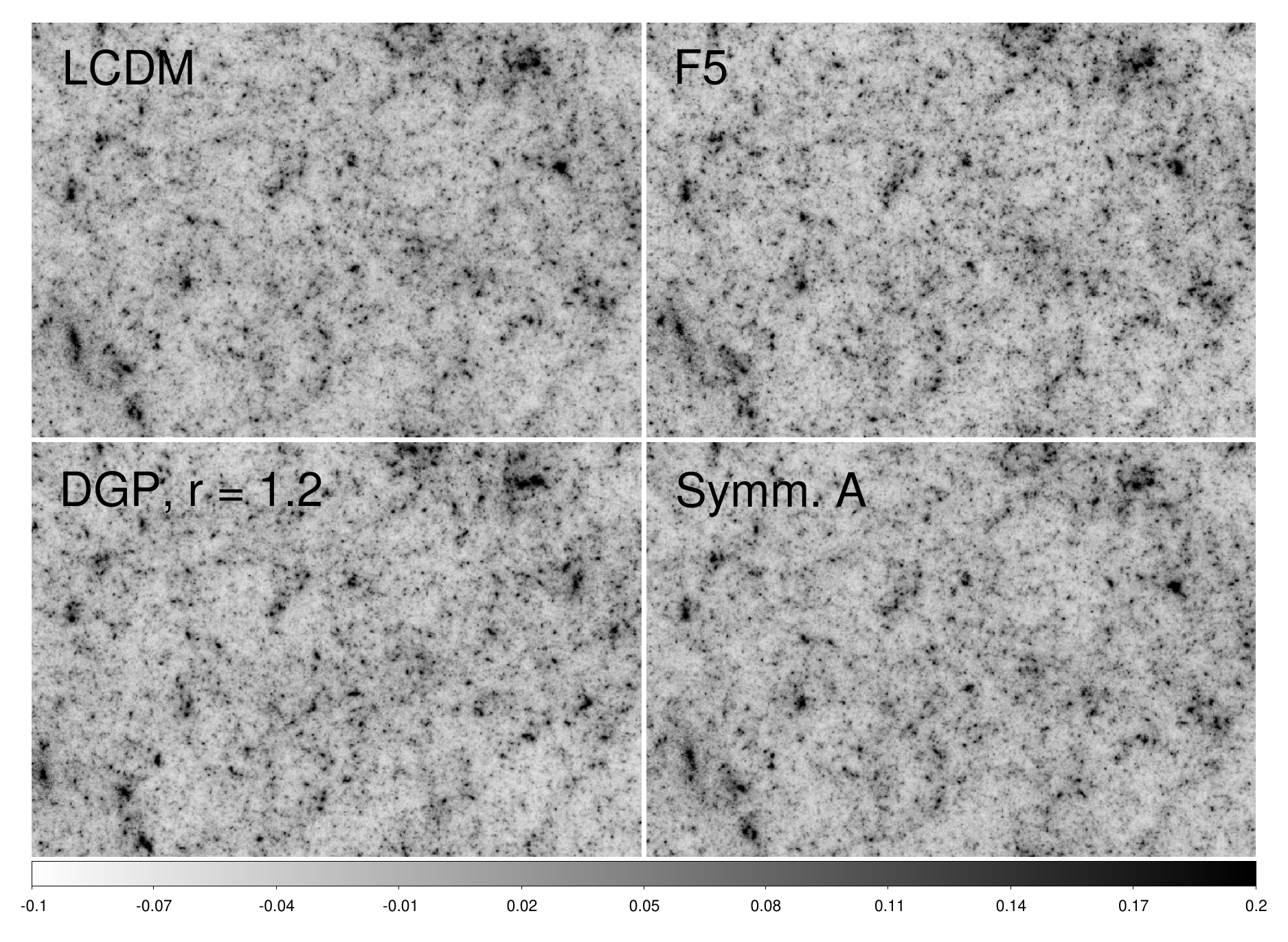}%
\caption{%
Comparison of a $1.85\deg \times 1.25\deg$ field of view of the convergence $\kappa$ in \LCDM\ and the modified gravity models, using a source redshift $z_s = 2$.
All four panels show the same light cone extracted from the respective simulations.
The convergence maps were simulated with the \code{GLAMER} pipeline for gravitational lensing.
}%
\label{fig:kappa}%
\end{figure}

Extracting lensing quantities from simulations using equations \eqref{eq:psi_cone} and \eqref{eq:kappa_cone} is cumbersome and numerically intensive, as each calculation and each desired source redshift involves raytracing through the simulation volume.
Instead, we simulate lensing by large-scale structure using the multi-plane approach of the \code{GLAMER} lensing pipeline \cite{2014MNRAS.445.1942M,2014MNRAS.445.1954P}.
An observed light cone is first segmented in the radial direction into a number of slices, and the three-dimensional mass distribution in each slice is projected onto a plane at the mean comoving distance of that slice.
The resulting two-dimensional surface mass density maps~$\kappa_i$, $i = 1, 2, \dots$ serve as lensing planes for \code{GLAMER}, which traces the propagation of light from plane to plane using the deflection angle
\begin{equation}
	\vec\alpha_i = \nabla\psi_i
\end{equation}
and the Poisson equation \cite{2014MNRAS.445.1954P}
\begin{equation}\label{eq:pois}
    \nabla^2 \psi_i = 2 \kappa_i
\end{equation}
which relates the surface mass density $\kappa_i$ to the Laplacian of the deflection potential $\nabla^2\psi_i$ on each plane.
Having thus constructed the lensing simulation, we are free to place a source plane at any redshift inside the light cone (i.e.\@ a delta distribution of source redshifts), and calculate maps of lensing quantities such as the convergence $\kappa$ or the shear $\gamma$ for an observer at redshift zero.
Sample convergence maps for the \LCDM, $f(R)$, symmetron, and DGP simulations are shown in figure~\ref{fig:kappa}.

The construction of light cones from simulations and the projection of the mass distribution onto individual planes is done by the \code{MapSim} tool \cite{2015MNRAS.452.2757G} in a single step.
Each light cone is constructed up to redshift $z_{\text{max}} = 2.0$, which is thus the highest source redshift available for the lensing maps.
The field of view of the light cone is a square with a side length of $3.85\deg$, which is the angle subtended by the simulation box size~$B_0$ at redshift $z_{\text{max}}$, giving us a total area of $14.82\deg^2$.

\begin{figure}%
\centering%
\includegraphics[width=\textwidth]{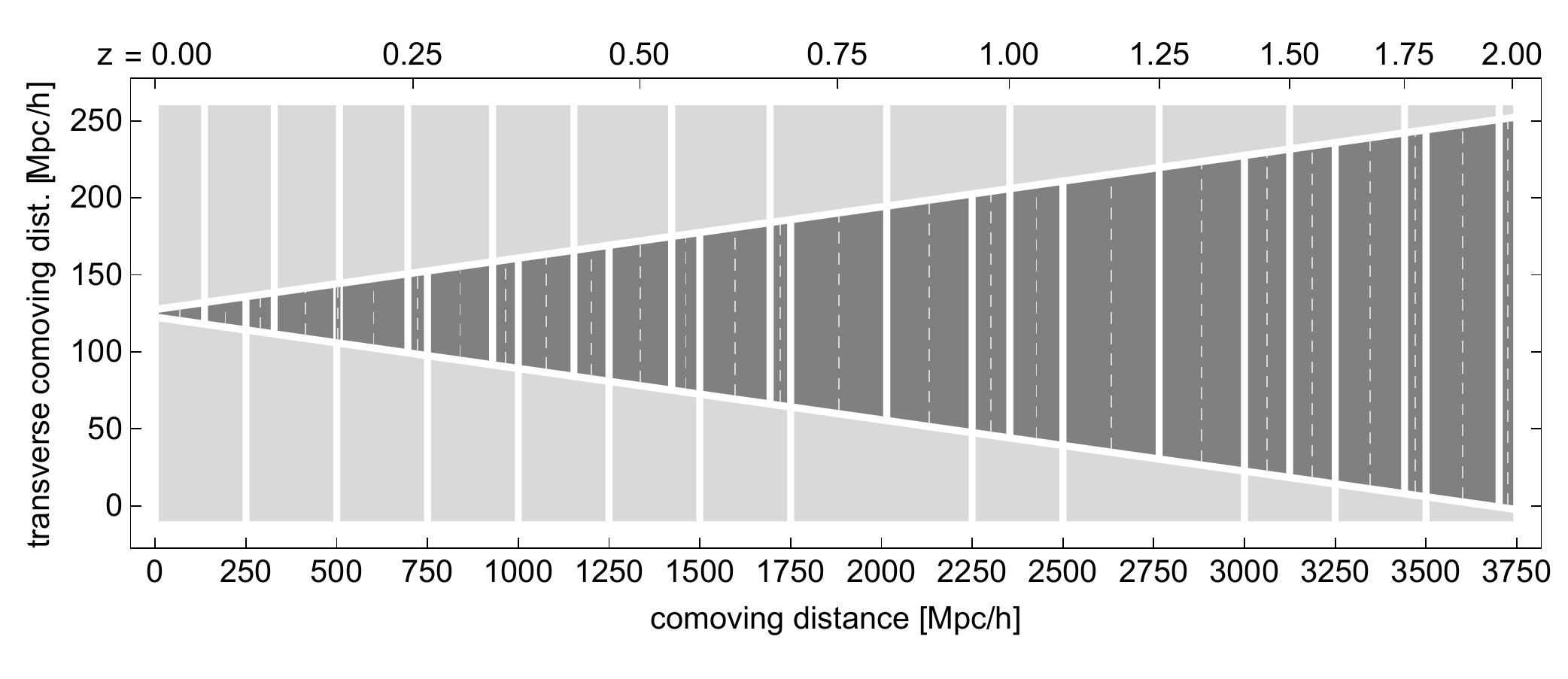}%
\caption{%
Light cone construction with \code{MapSim} for the \LCDM\ simulation.
The simulation box size is $B_0 = 250~\Mpc/h$.
The dark shaded area is the light cone, growing to a transverse comoving size equal to the box size $B_0$.
The mass distribution in each of the individual light cone segments is projected onto a lensing plane at the centre of the segment (dashed line).
The regions above the light cone indicate the simulation snapshots used to construct each segment.
The regions below the light cone indicate groups of segments that have been randomised in the same way.
}%
\label{fig:mapsim}%
\end{figure}

Figure~\ref{fig:mapsim} shows the schematic construction of a light cone for the \LCDM\ simulation.
It is clear that for lower redshifts, much of the simulation box volume is unused.
Randomisation of the box volume through rotation and translation offers a way to extract multiple light cones from a single simulation \cite{2000ApJ...530..547J}, where each of the light cones ends up with a random portion of the simulation box.

\code{MapSim} constructs a light cone by randomly picking an origin and orientation of the box volume, defining zero comoving distance and the direction of the line of sight.
Increasing the comoving distance, particles in the field of view are mapped into the light cone, making use of the periodic boundary conditions of the simulation box.
This continues until the comoving distance is a multiple of the simulation box size $B_0$, changing snapshots as they become a better fit for current redshift.
The process is repeated, starting from the randomisation, until the whole light cone has been constructed.

Using the technique laid out above, we extract ten randomised light cones from each of the simulations and simulate the gravitational lensing of the contained large scale structure.
We then create lensing maps of the convergence fields $\kappa$ for source redshifts $z_S = 0.5$, $1.0$, $1.5$, $2.0$.
These maps have a size of $2048 \times 2048$ pixels for the aforementioned field of view of $3.85\deg \times 3.85\deg$, resulting in an angular resolution of $6.77\ \text{arcsec}$ per pixel.
We note that since the simulation box fills the whole field of view at $z = 2.0$, the same large-scale structure is present in each light cone for the higher redshift slices.
We thus expect to underestimate the sample variance with increasing redshift.


\section{Results}
\label{results}

We now discuss our results, briefly focusing on the power spectrum of the density field before we turn to the convergence power spectrum and its evolution.


\subsection{Matter power spectrum}

\begin{figure}[tbp]%
\begin{tabular}{ccc}%
\scriptsize\hspace{3.2em} $F_5$ &
\hfill&
\scriptsize\hspace{3.2em} $F_6$
\\[-1.2pt]
\includegraphics[width=.42\textwidth]{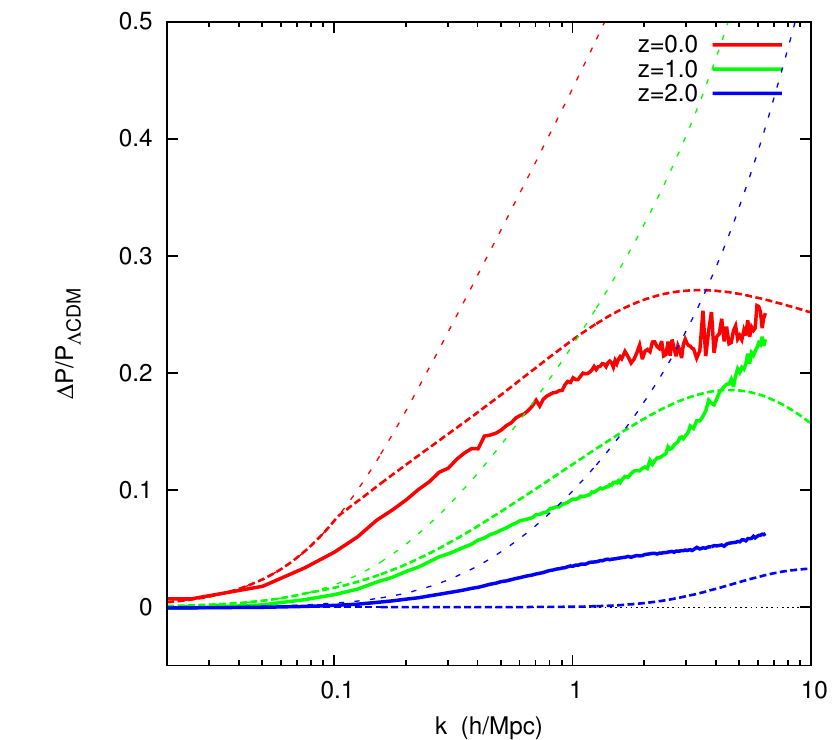} &
&
\includegraphics[width=.42\textwidth]{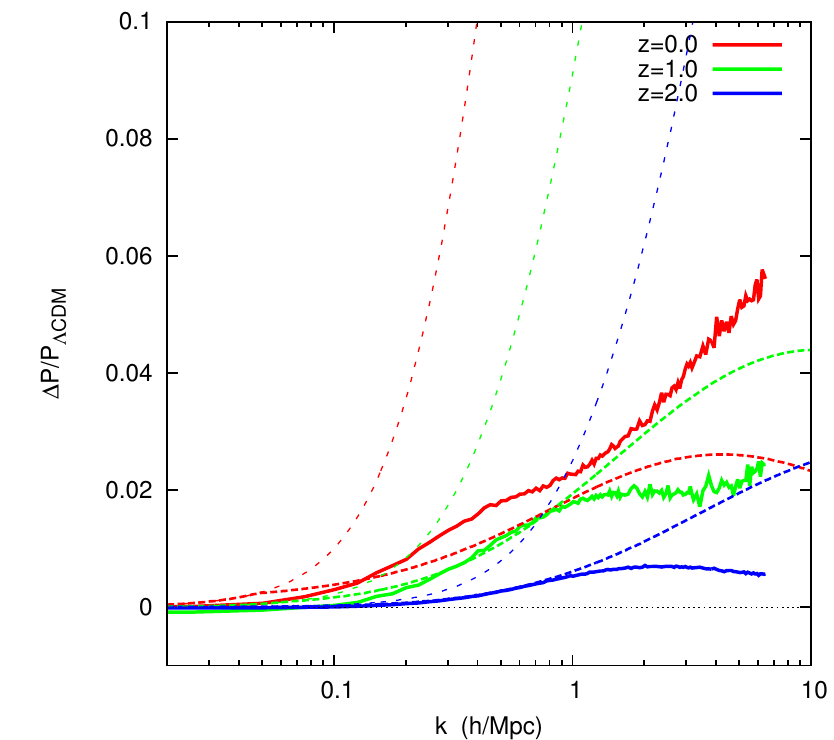}
\\[5pt]
\scriptsize\hspace{3.2em} Symm.~$A$ &
&
\scriptsize\hspace{3.2em} Symm.~$B$
\\[-1.2pt]
\includegraphics[width=.42\textwidth]{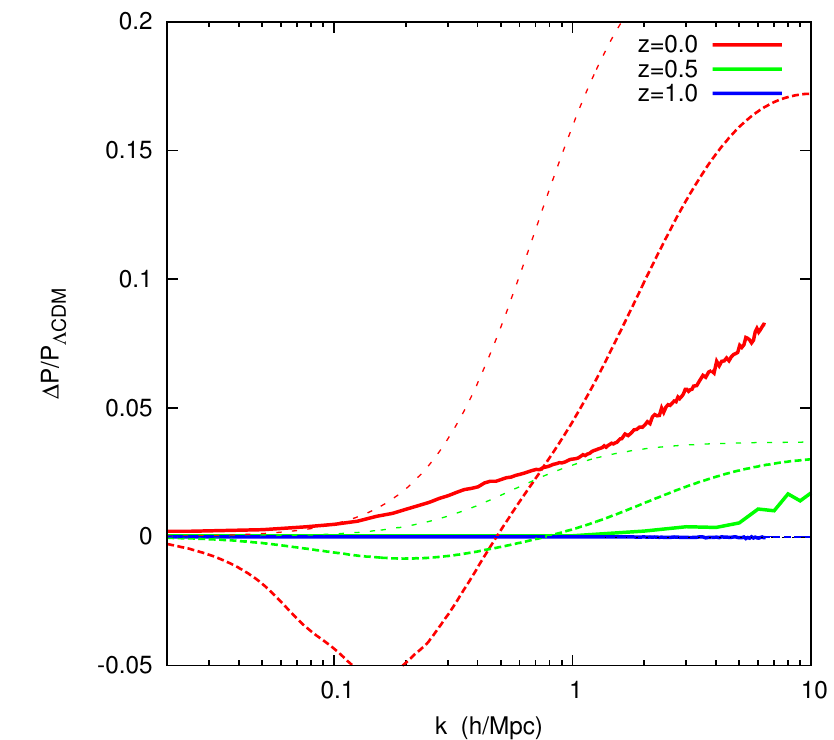} &
&
\includegraphics[width=.42\textwidth]{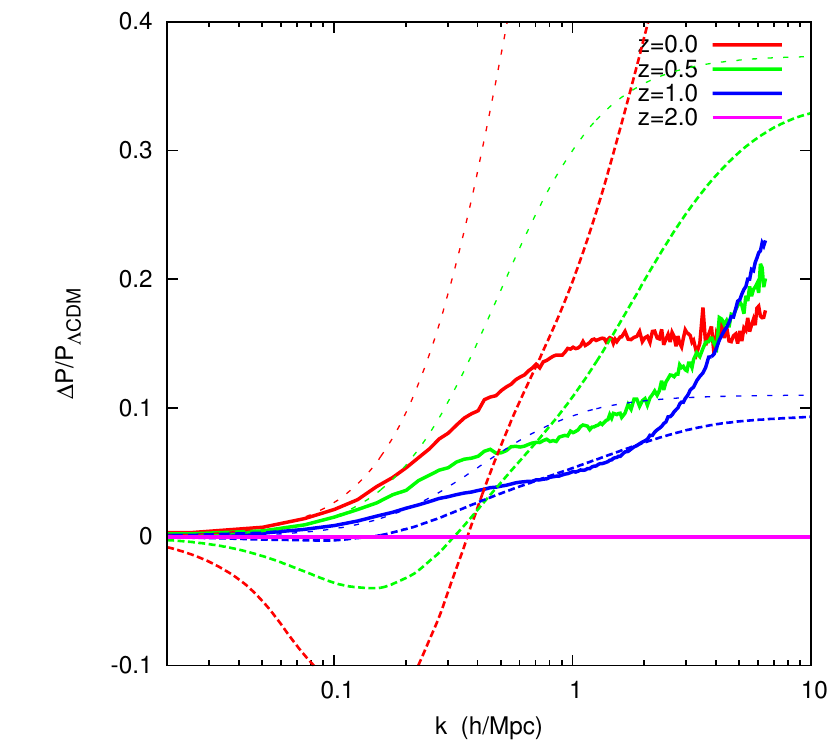}
\\[5pt]
\scriptsize\hspace{3.2em} DGP, $r = 1.2$ &
&
\scriptsize\hspace{3.2em} DGP, $r = 5.6$
\\[-1.2pt]
\includegraphics[width=.42\textwidth]{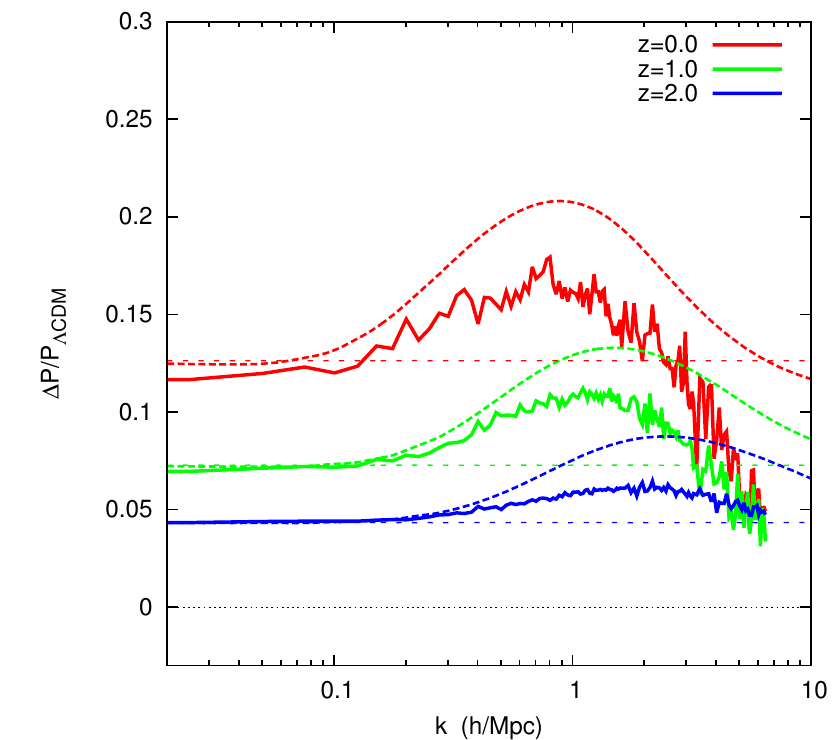} &
&
\includegraphics[width=.42\textwidth]{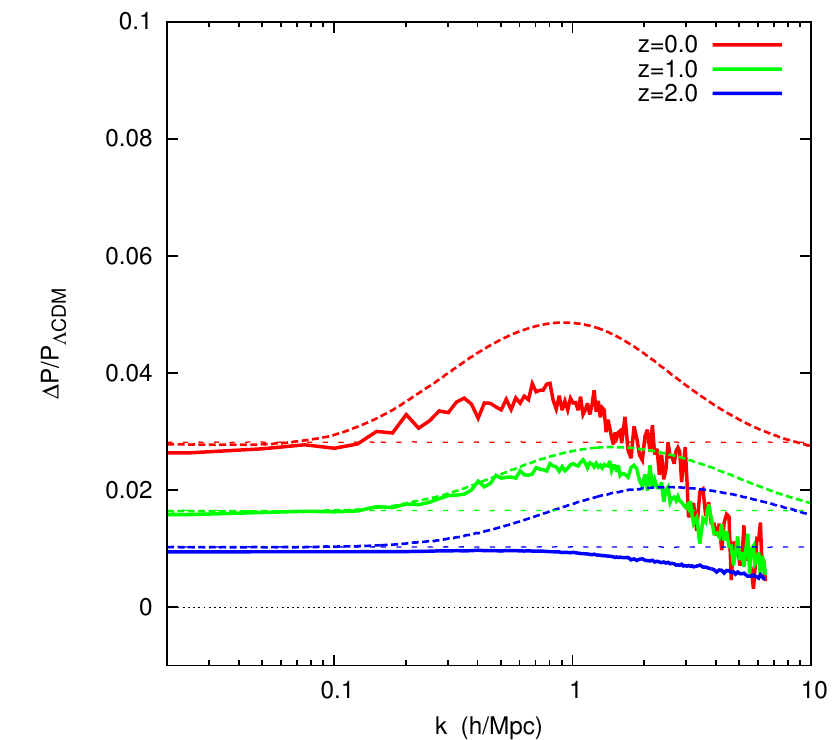}
\end{tabular}%
\caption{%
The fractional difference in the matter power spectrum with respect to \LCDM\ for the $f(R)$, symmetron, and DGP models.
The thin dotted lines shows the linear perturbation theory and the dashed lines shows the \code{halofit} predictions (using the \code{halofit} fitting function of \code{MGCAMB} for $f(R)$).}%
\label{fig:pofk}%
\end{figure}

The matter power spectrum is defined via
\begin{equation}\label{eq:P_k}
\left<\delta_m(\vec{k})\delta_m(\vec{q})\right> = (2\pi)^3\delta^{(3)}(\vec{k}+\vec{q})P(k),
\end{equation}
where $\delta_m(\vec{k})$ is Fourier transform of the density field $\delta_m(\vec{x}) \equiv \rho_m(\vec{x})/\overline{\rho}_m - 1$.
We have calculated the matter power spectrum for all of our simulations using \code{POWMES} \cite{2011ascl.soft10017C}.
Our obtained power spectra agree well with previous $N$-body simulations of the same models \citep{2012JCAP...01..051L,2014A&A...562A..78L,2013JCAP...05..023L,2013MNRAS.436..348P,2008PhRvD..78l3524O,2012ApJ...748...61D,2012JCAP...10..002B}, which serves as a good cross-check for the validity of our results.
We have also computed both the linear predictions, using Eq.~\eqref{eq:lineq}, and the \code{halofit} \cite{2003MNRAS.341.1311S} predictions for the non-linear power spectrum using \code{MGCAMB} \cite{2011JCAP...08..005H}.
This latter code is a modified gravity modification of the Boltzmann-code \code{CAMB} \cite{2000ApJ...538..473L} and uses the improved fitting formula (designed for \LCDM) of \cite{2012ApJ...761..152T} to get better agreement with simulations on small scales.
The fractional difference of the matter power spectrum with respect to \LCDM\ as a function of redshift can be seen in Fig.~\ref{fig:pofk} for the $f(R)$, symmetron, and DGP models, respectively. 

Before we discuss the results in more detail, we note that, overall, the results we find confirm the qualitative features we discussed in the theory section.
For $f(R)$ and the symmetron model, we see that the power spectrum approaches \LCDM\ on large scales, while for DGP there are modifications on all scales.
Furthermore, comparing the full simulation results with the predictions of linear theory shows the effect of screening: linear theory greatly overpredicts the amount of clustering on small scales.
This confirms the point we have made above: non-linear effects are crucial for accurately modelling the effects of gravitational screening and methods, such as the ones presented in this paper, play an essential role.
The \code{halofit} predictions will be discussed in more detail below.

\paragraph{\boldmath $f(R)$}

For this model, modifications to gravity boost structure formation on small scales.
Furthermore, we find that $P(k) \simeq P(k)_{\Lambda\rm CDM}$ for scales $k\lesssim 0.05~h/\Mpc$ and over all times.
The effects are stronger in the $F_5$ model than in the $F_6$ model and this is due to the larger range (see Eq.~\eqref{eq:mbetafofr}) of the fifth-force in the former simulation.
For the $F_5$ ($F_6$) model we see that the matter power spectrum is enhanced by up to $\approx 25\%$ ($\approx 2\mbox{--}5\%$) at $z=0$ for non-linear scales $k\sim 1 \mbox{--} 10~h/\Mpc$.
At $z=2$ the deviations from \LCDM\ are below $5\%$ and $1\%$ for all scales in the $F_5$ and $F_6$ model respectively.
At earlier times the modifications are even smaller and thus, for these models, modifications to gravity only have a potentially observable impact on structure formation at late times, $z\lesssim 2$.
The main reason for this is that the range of the scalar field (again see Eq.~\eqref{eq:mbetafofr}) decreases rapidly as we go to high redshift.
Comparing the results of linear perturbation theory with our simulation results, we see that the amount of clustering is greatly overpredicted in linear theory.
For example, at $k = 1~h/\Mpc$ and $z=0$, linear theory predicts a modification of $\approx 50\%$ in the $F_5$ model whereas the actual result is closer to $\approx 20\%$.

\paragraph{Symmetron}

Modifications of gravity in the symmetron model follow the same pattern as in $f(R)$ gravity, where structure formation is boosted on small scales.
The effects are stronger in the $B$ model than in the $A$ model.
The reason for this is the smaller value of $a_{\rm ssb}$ in the former simulation, which means that the fifth-force has been active for a longer period of cosmic time.
The fifth-force is not in operation before the time of symmetry breaking $a = a_{\rm ssb}$ and consequently the power spectrum is the same as in \LCDM\ for earlier times $a<a_{\rm ssb}$.
For the $A$ ($B$) model we see that the matter power spectrum is enhanced by up to $\approx 15\%$ ($\approx 5\%$) at $z=0$ for non-linear scales $k\sim 1 \mbox{--} 10~h/\Mpc$.
Linear theory is an even worse fit to the simulation results in the symmetron model than in $f(R)$ gravity.
At $k=1~h/\Mpc$ at $z=0$ we find a $\approx 2\%$ modification for the $B$ model whereas linear theory predicts almost $20\%$.
This larger deviation can be attributed to the stronger screening mechanism in the symmetron model, i.e.\@ the fact that $\beta \propto \phi$ leads to additional screening on small scales (and high density regions) where $\phi$ is clustered close to $\phi = 0$.

\paragraph{DGP}

Contrary to the two models discussed above, gravity is modified on all scales in the DGP model.
At $z=0$ we find that structure formation in the $r_c H_0 = 1.2$ ($=5.6$) model is enhanced by $\approx 12\%$ ($\approx 3\%$).
Going to earlier times the modifications become smaller and, at $z=2$, are less than $\approx 4\%$ and $\approx 1\%$ in the two models, respectively.
Since $\beta_{\rm DGP}$ increases with increasing $r_c$ we expect that a larger $r_c$ leads to stronger modifications, and this is indeed the case in our results.
On highly non-linear scales $k\gtrsim 5~h/\Mpc$, deviations in the power spectrum are seen to drop for both simulations and at $k = 10~h/\Mpc$ we are very close to the \LCDM\ prediction.
This is due to the Vainshtein mechanism being in play and reducing the effects of the fifth-force on the small-scale structure formation.
We also see that for the two models we have chosen to simulate, the relative difference $\Delta P/P_{\Lambda\rm CDM}$ has a similar shape at all three redshifts depicted in Fig.~\ref{fig:pofk}, while the amplitude is markedly different.


\subsection{Convergence power spectrum}
\label{sec:C_l-absolute}

Similar to the power spectrum~\eqref{eq:P_k}, the convergence power spectrum $C_\ell$ is formally defined by
\begin{equation}\label{eq:C_l}
	\left\langle \hat\kappa(\vec\ell) \hat\kappa(\vec\ell') \right\rangle = (2\pi)^2 \, \delta^{(2)}(\vec\ell + \vec\ell') \, C_\ell \;,
\end{equation}
where $\ell$ denotes the angular mode.
The convergence power spectrum is an important cosmological probe, as it can be related to the evolution history of the universe via Limber's approximation through
\begin{equation}\label{eq:clzs}
	C_l(z_s) = \frac{9\Omega_{m}^2H_0^4}{4}\int_0^{z_s}dz \frac{g^2(z)(1+z)^2}{\chi^2(z)H(z)}P(k = l/r(z), z)
\end{equation}
where $P(k,z)$ is the matter power spectrum at redshift~$z$ and $g(z)$ is the lensing weight.
For a single source plane at $z=z_s$ we have $g(z) = \chi(z)(\chi(z) - \chi(z_s))/\chi(z_s)$ where $\chi(z)$ is the co-moving distance.
This is a powerful combination of many cosmological quantities which can help break degeneracies arising in other probes.
The convergence power spectrum is also directly related to all observed quadratic statistics of cosmic shear measurements, and might soon be measured directly (see e.g.\@ \cite{2012MNRAS.426.1262H,2015RPPh...78h6901K}).

\begin{figure}%
\includegraphics[width=.49\textwidth]{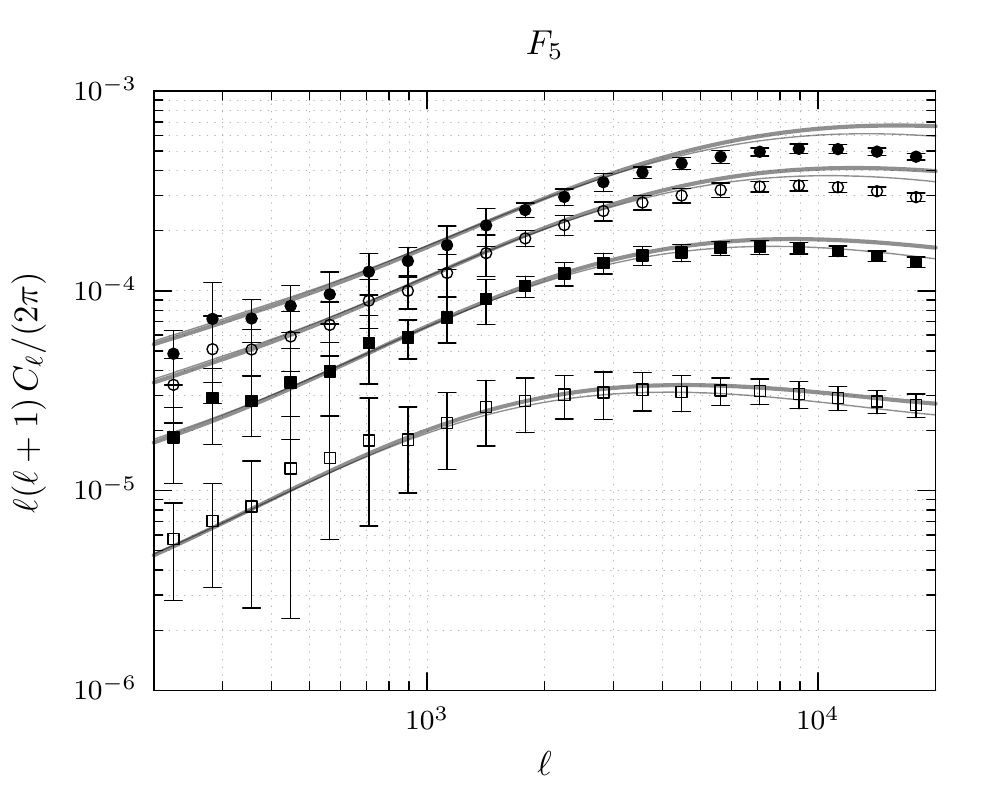}%
\hfill%
\includegraphics[width=.49\textwidth]{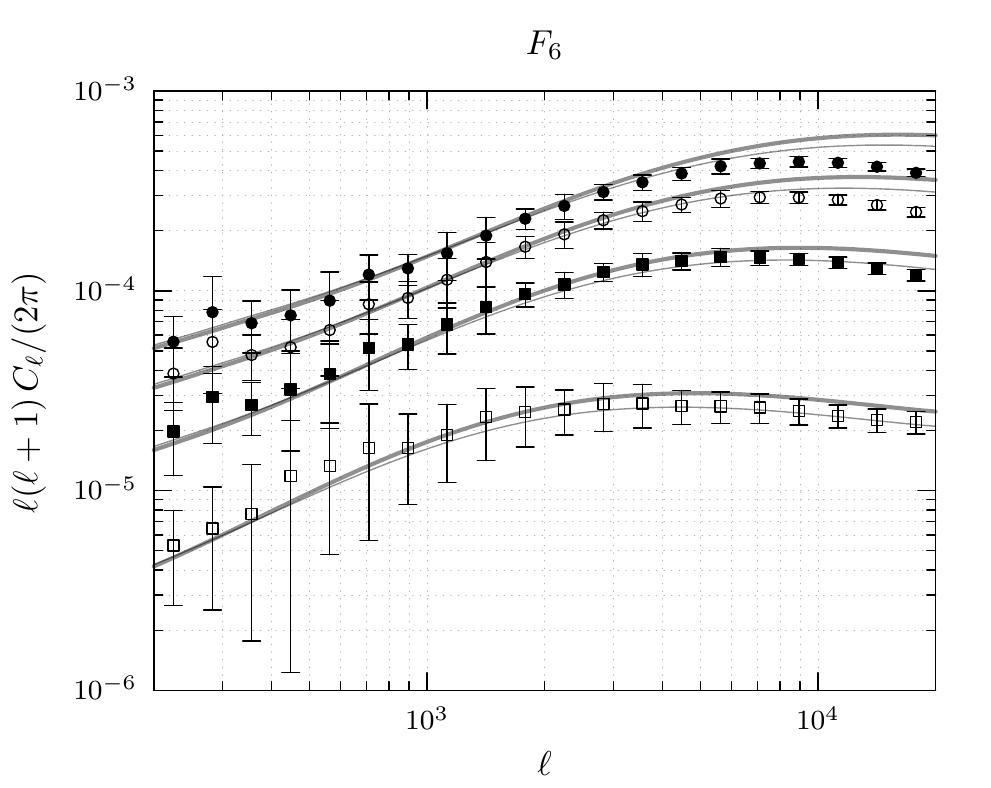}%
\\[\baselineskip]%
\includegraphics[width=.49\textwidth]{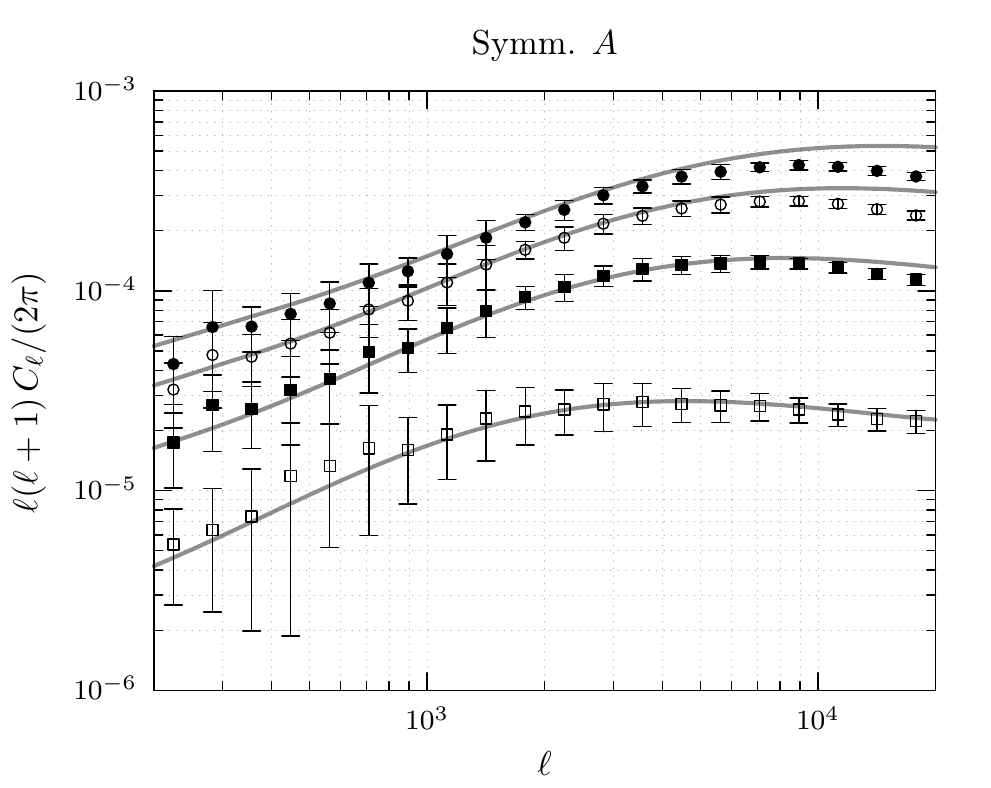}%
\hfill%
\includegraphics[width=.49\textwidth]{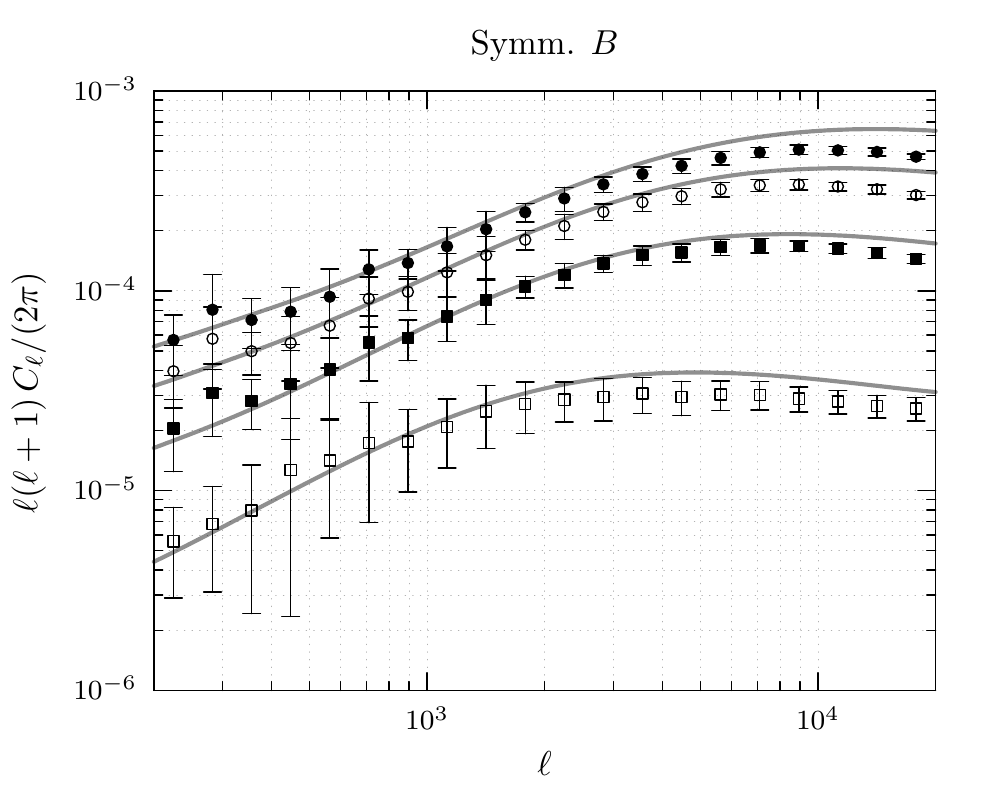}%
\\[\baselineskip]%
\includegraphics[width=.49\textwidth]{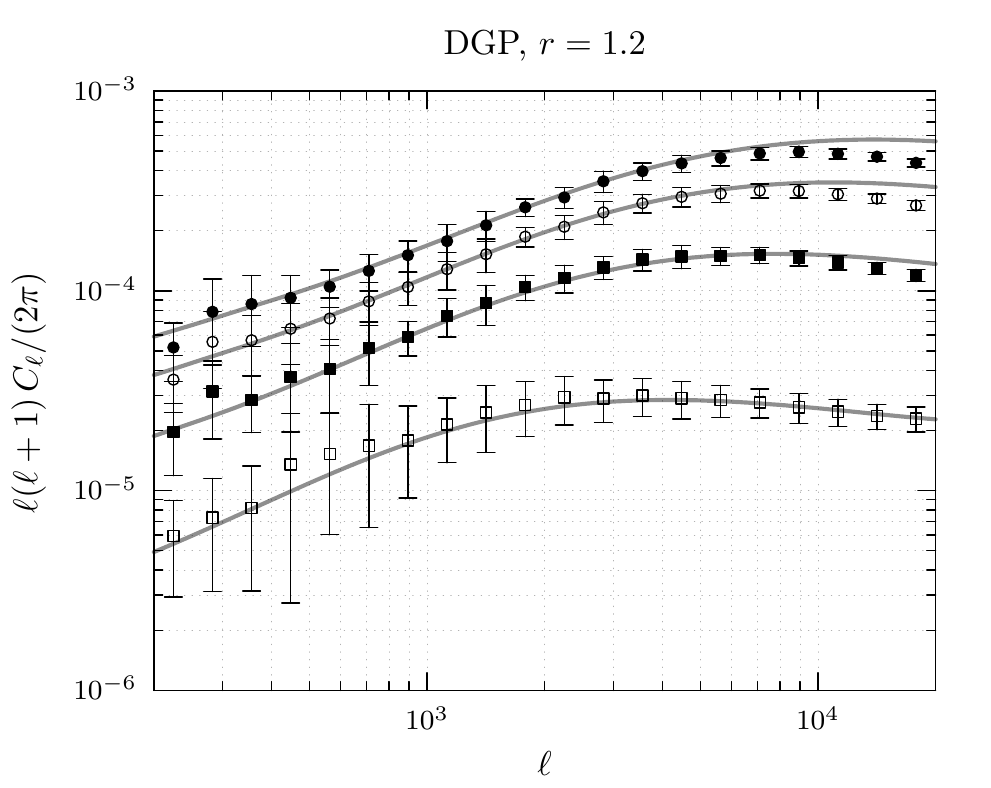}%
\hfill%
\includegraphics[width=.49\textwidth]{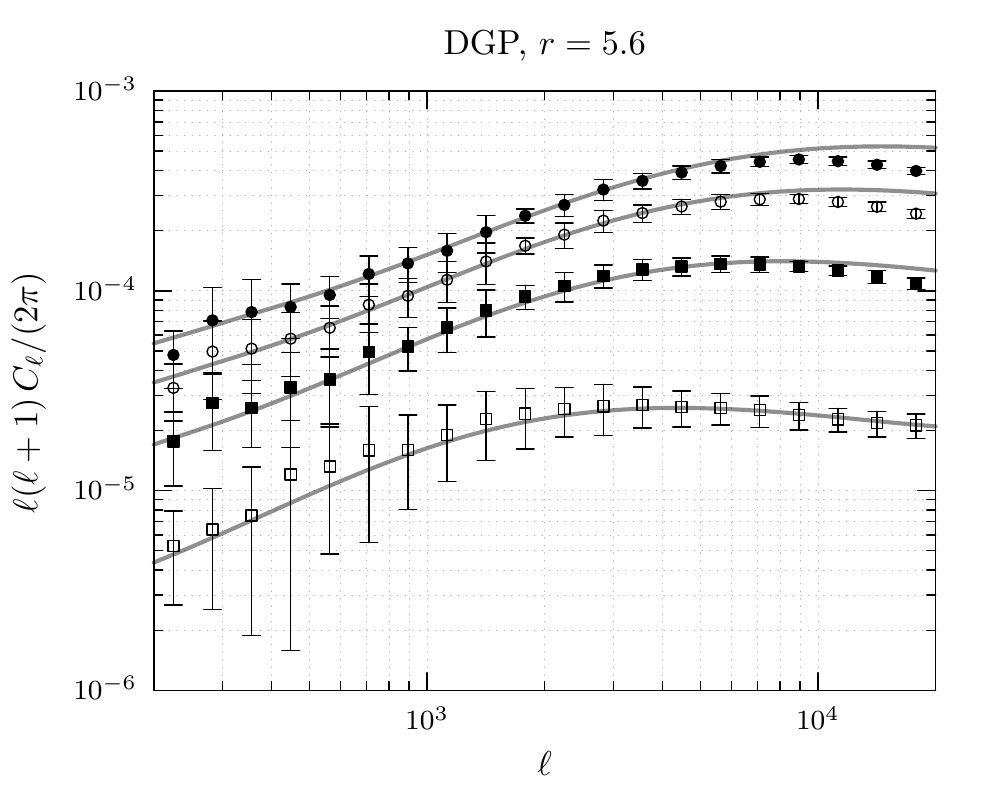}%
\caption{%
Dimensionless convergence power spectra $\ell (\ell+1) \, C_{\ell}$ for the $f(R)$, symmetron, and DGP simulations at source plane redshifts $z_S = 0.5, 1.0, 1.5, 2.0$ (bottom to top).
Shown are the $1\sigma$ error bars over the sample of realisations.
Also shown are the predictions obtained from \code{halofit} (thick) and, in the case of $f(R)$, \code{mghalofit} (thin).
}%
\label{fig:cl}%
\end{figure}

In order to estimate the power spectrum from our simulated convergence maps, we use
\begin{equation}
	C_\ell = \frac{1}{A} \left\langle \left| \hat\kappa(\vec\ell) \right|^2 \right\rangle \;,
\end{equation}
where $\hat\kappa$ is the Fourier transform of the convergence field, and the averaging is done over the angle of vector $\vec\ell$.
In practice, we calculate $\hat\kappa$ via a Fast Fourier Transform of the convergence maps (after zero-padding to mitigate boundary effects), and perform the averaging in bins of $\Delta\ell$ with logarithmic spacing.
By calculating the convergence power spectrum separately for each simulated light cone, we arrive at a sample of results which further gives us a handle on the sample variance of our results.
The estimated power spectra for our respective $f(R)$, symmetron, and DGP models are shown in Fig.~\ref{fig:cl}.

\begin{figure}%
\centering%
\includegraphics[width=.66\textwidth]{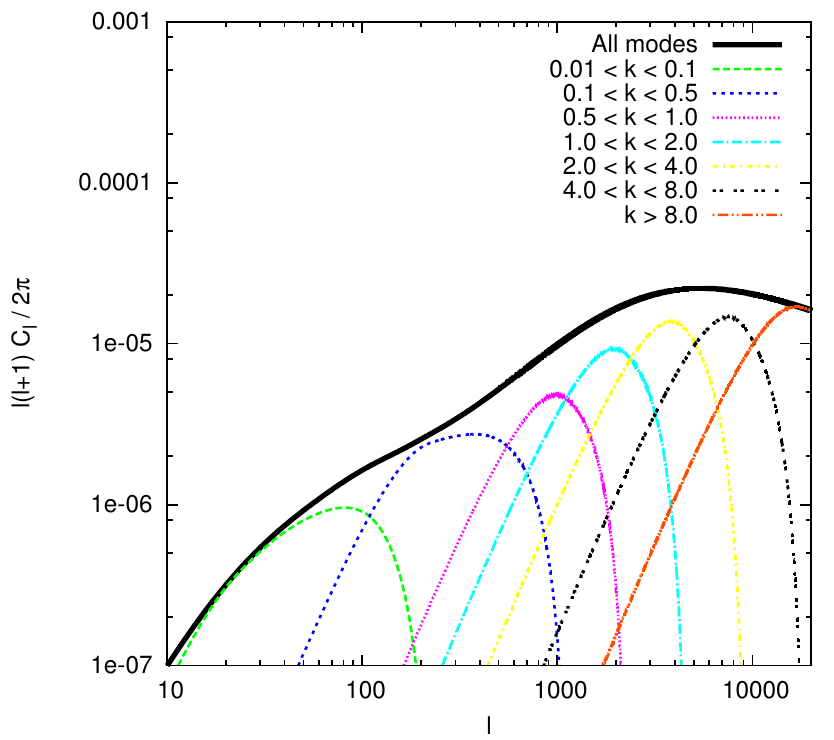}%
\caption{%
We show the relative contribution to the convergence power spectrum from different ranges of co-moving Fourier modes of the density fluctuations.
The plot is made by computing the \code{halofit} prediction for $C_\ell$ and taking $P(k,z) = 0$ for $k$ values outside the range indicated in the plot. The source is here at $z=1.0$.
}%
\label{fig:cl_theta_modes}%
\end{figure}

To estimate up to which $\ell$ we can trust our results, we computed the relative contribution to $C_\ell$ coming from different ranges of co-moving modes of the matter power spectrum.
The result can be seen in Fig.~\ref{fig:cl_theta_modes}.
The simulations we have performed have a particle Nyquist frequency of $k_{\rm Ny} \simeq 6.5~h/\Mpc$.
Fixed-grid simulations with different box sizes usually start to deviate from each other for modes larger than $k\sim k_{\rm Ny}/4 \mbox{--} k_{\rm Ny}/2$, resulting in $k_{\rm max} \sim 2 \mbox{--} 3~h/\Mpc$ and $\ell_{\rm max} \sim 2000 \mbox{--} 3000$.
However, our simulations have adaptive refinements, which means that the effective Nyquist frequency is much larger and a rough estimate from the refinement structure gives us a factor 5--10 at $z=0.0$.
From this we estimate that we can trust the $C_\ell$ spectra up $\ell_{\rm max}\sim 10^4$.
For our largest source redshift $z_s = 2.0$ this $\ell_{\rm max}$ value is probably too large due to the lack of refinement at early times.
The minimum $\ell$-value we can study is fixed by the simulation box size $B_0 = 250~\Mpc/h$, corresponding to values of $k_{\rm min} = 0.025~h/\Mpc$ and $\ell_{\rm min} \sim 100$ (see Fig.~\ref{fig:cl_theta_modes}).


\subsection{Evolution of the convergence power spectrum}
\label{sec:C_l-relative}

\begin{figure}%
\includegraphics[width=.49\textwidth]{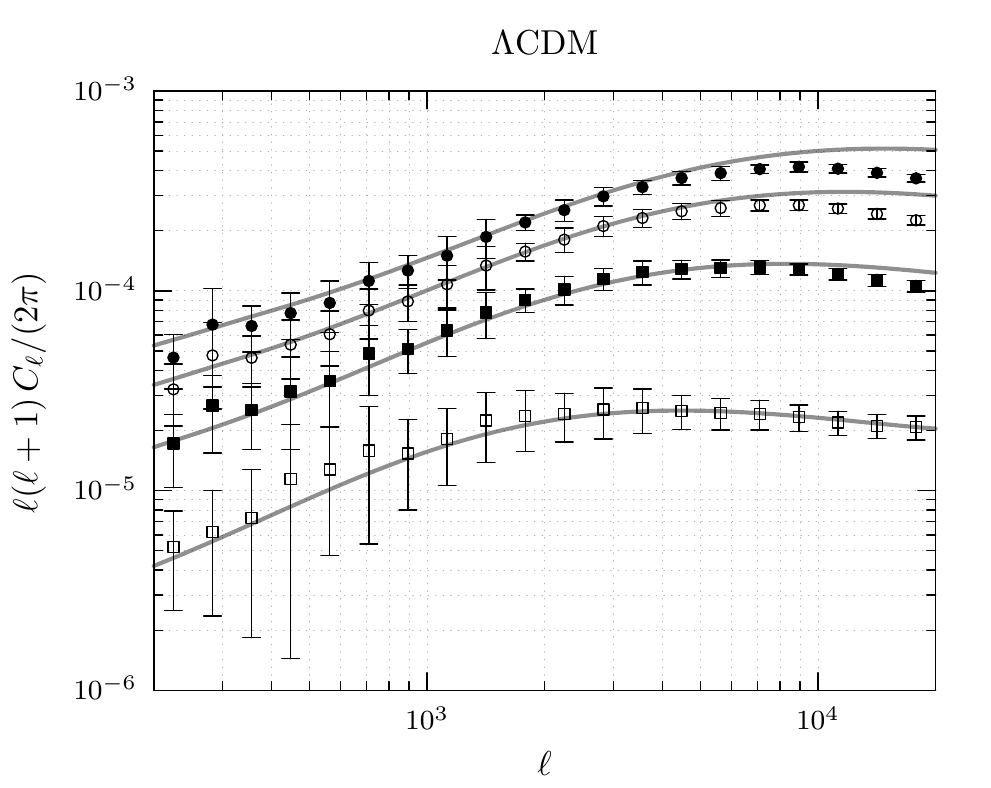}%
\hfill%
\includegraphics[width=.49\textwidth]{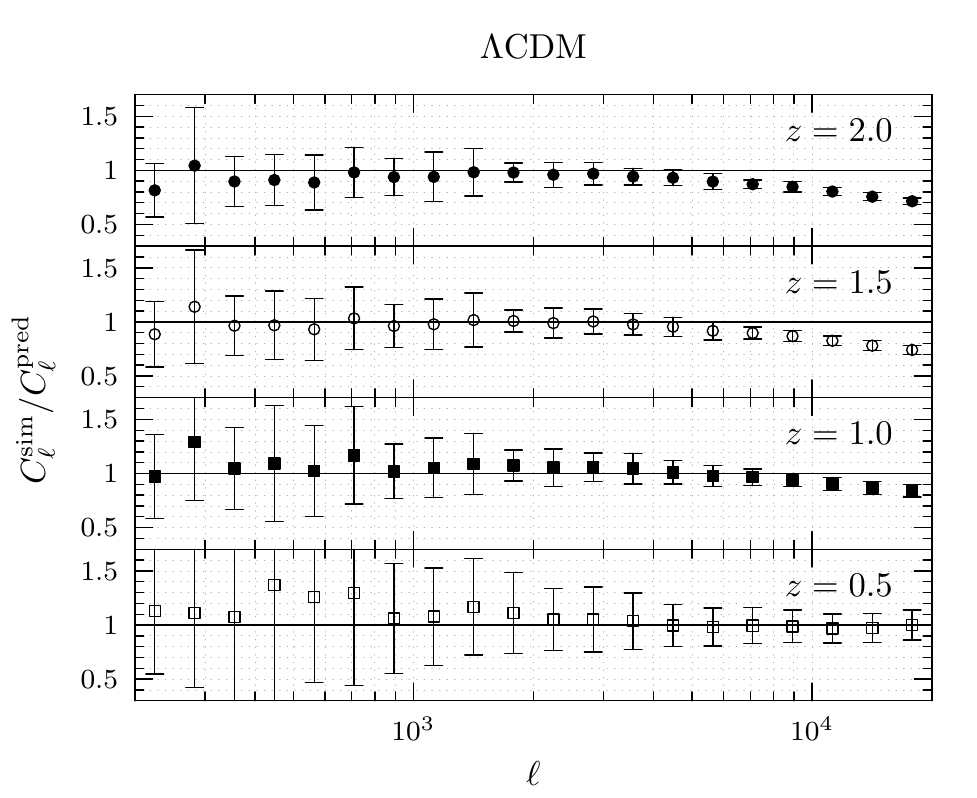}%
\caption{%
\emph{Left:}\ Dimensionless convergence power spectrum $\ell (\ell+1) \, C_{\ell}$ for the \LCDM\ simulation at source plane redshifts $z_S = 0.5, 1.0, 1.5, 2.0$ (bottom to top).
Shown are the $1\sigma$ error bars over the sample of realisations.
Also shown is the prediction obtained from \code{halofit}.
\emph{Right:}\ Difference between the \LCDM\ simulation and predictions from \code{CAMB}.
}%
\label{fig:cl_lcdm}%
\end{figure}

We will first investigate the redshift evolution of the non-linear power spectrum\footnote{%
See \cite{2008PhRvD..78d3002S} for a discussion on the effect of modified gravity on the weak lensing convergence power spectrum at linear scales.
}
in the screened models relative to a reference \LCDM\ simulation (see Fig.~\ref{fig:cl_lcdm}).
The reason for this is twofold:
First, as noted before, we expect a loss of power in the high $z$, high $\ell$ regime due to the lack of refinement of the simulations at early times.
However, this loss of power is consistent among the different simulations, which all evolved with the same refinement settings.
Although we are deep in the non-linear regime, and the effect of resolution may be different for different models, we expect that the power ratio $C_\ell/C_\ell^{\Lambda\mathrm{CDM}}$ with respect to \LCDM\ will be close to the true value over the full range up to $\ell_{\rm max} \sim 10^4$ and indicative of the differences one might find at higher resolution.
Another reason for investigating the power ratio is of a more physical nature.
If $C_\ell/C_\ell^{\Lambda\mathrm{CDM}}$ is constant in both redshift $z$ and mode $\ell$, it is in principle indistinguishable from \LCDM\ with a different normalisation of the power spectrum.
If $C_\ell/C_\ell^{\Lambda\mathrm{CDM}}$ only depends on $z$ (and not $\ell$), it might also be indistinguishable from \LCDM\ with a different background history.
Thus a robust signature of modified gravity requires an evolution in both redshift and scale.

\begin{figure}[tbp]
\includegraphics[width=.49\textwidth]{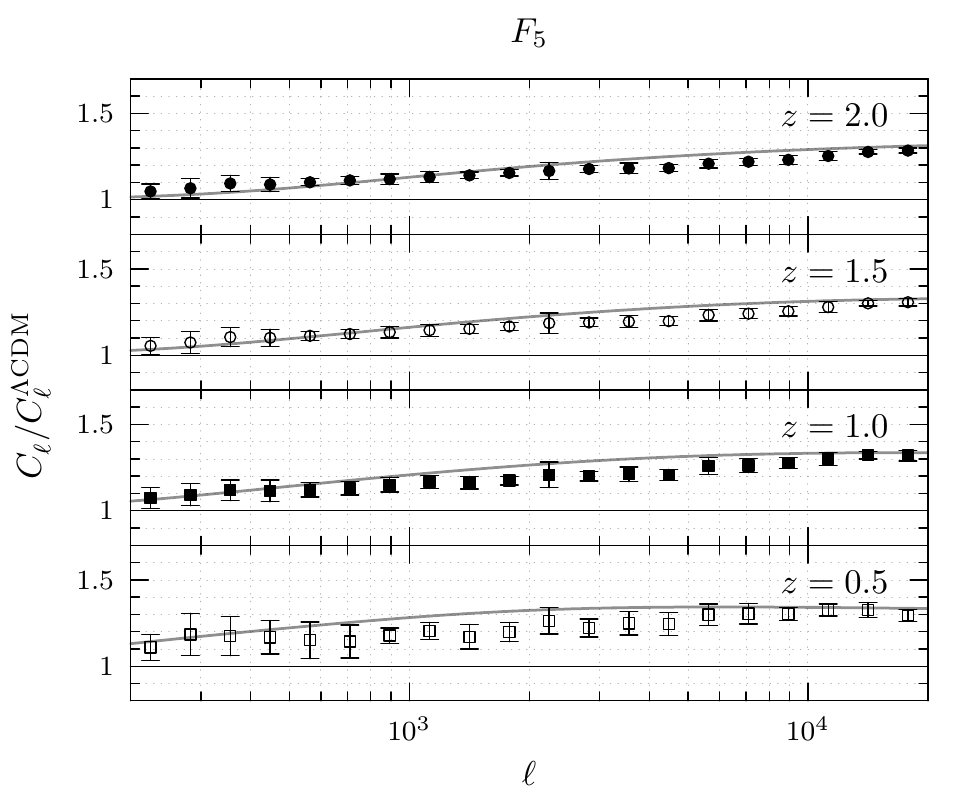}%
\hfill%
\includegraphics[width=.49\textwidth]{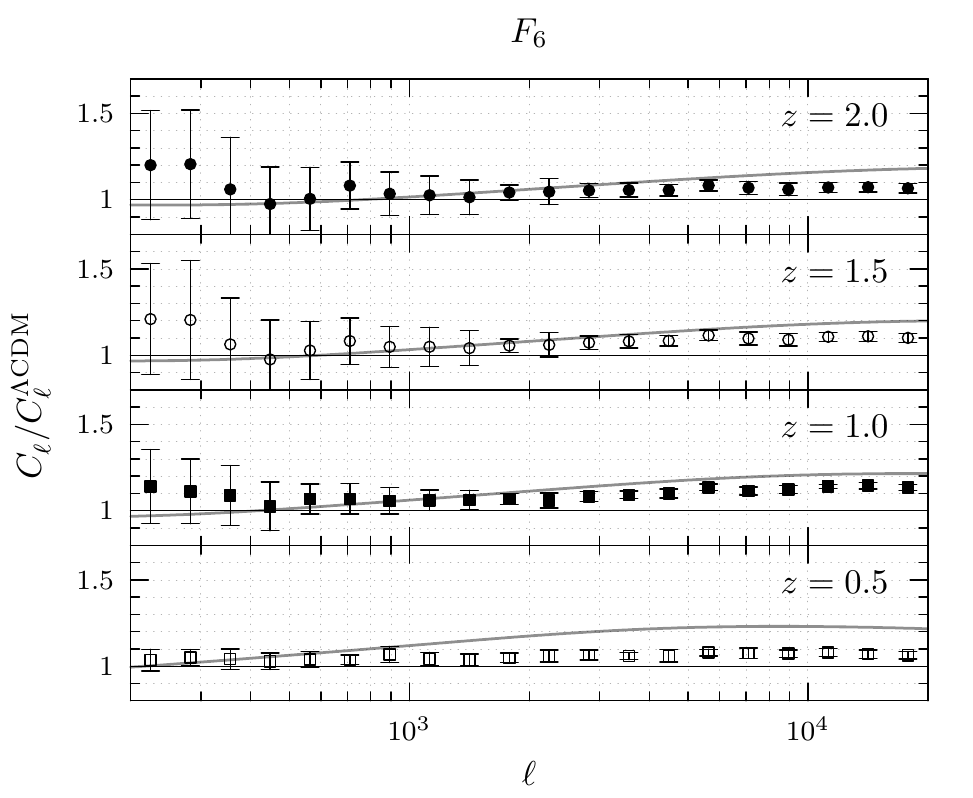}%
\\[\baselineskip]%
\includegraphics[width=.49\textwidth]{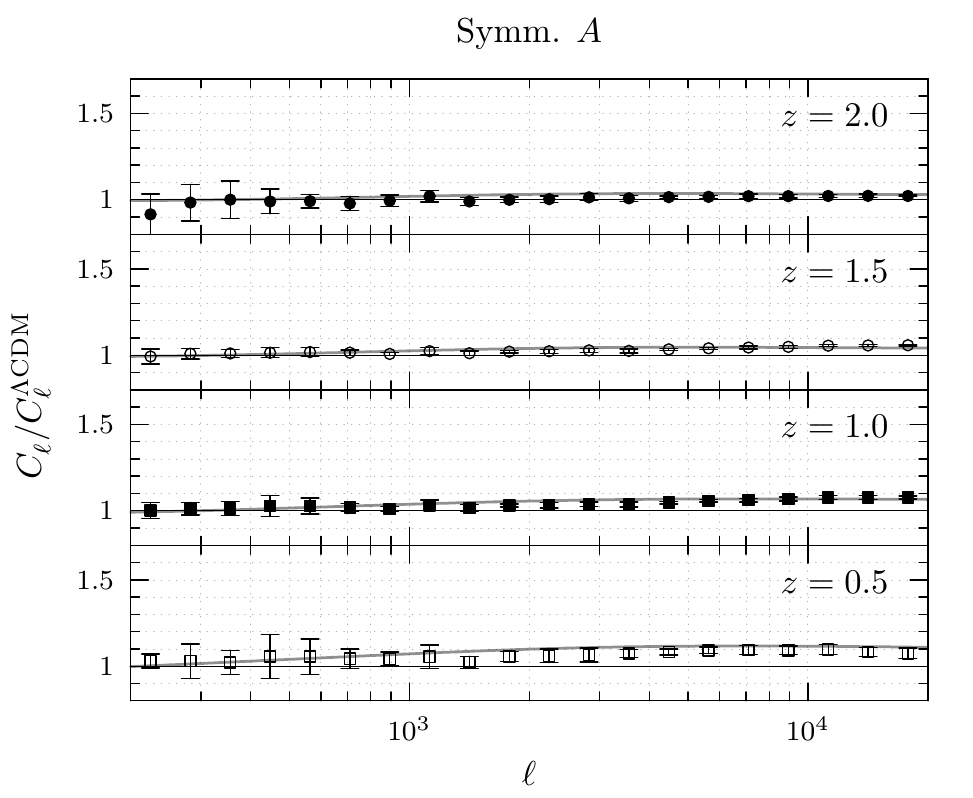}%
\hfill%
\includegraphics[width=.49\textwidth]{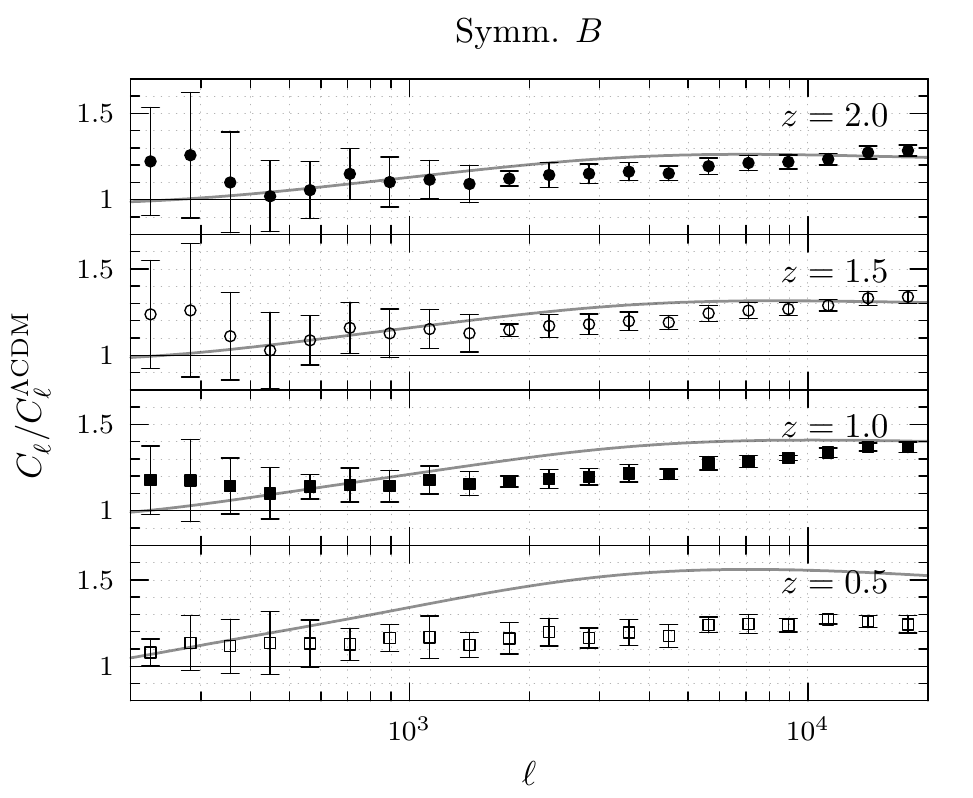}%
\\[\baselineskip]%
\includegraphics[width=.49\textwidth]{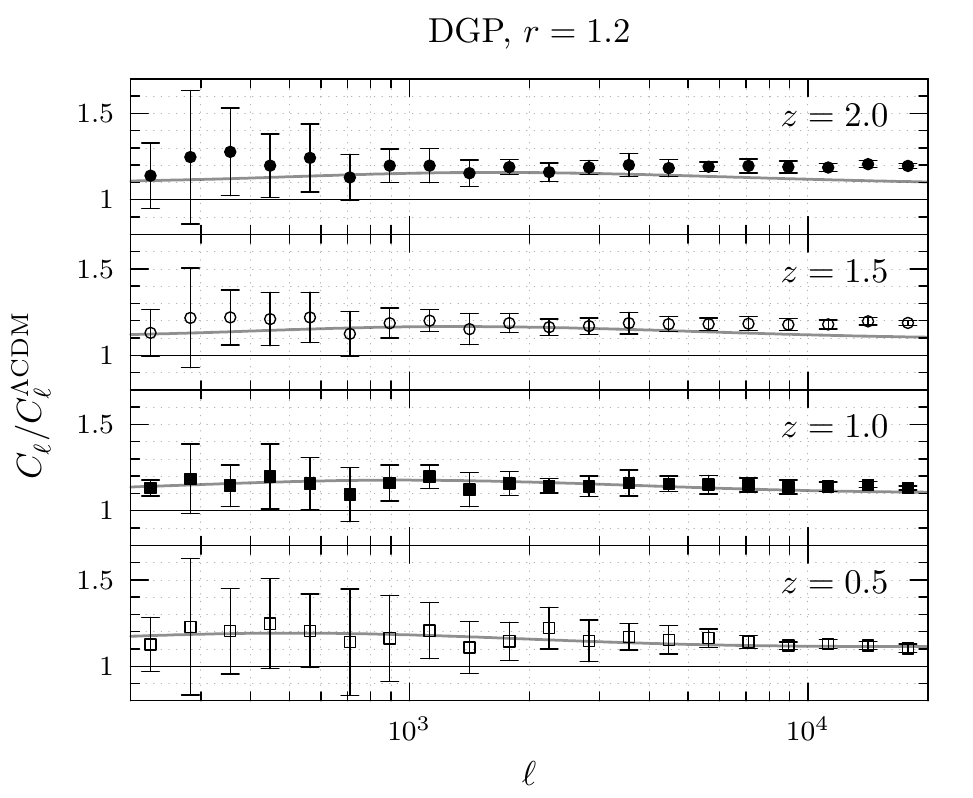}%
\hfill%
\includegraphics[width=.49\textwidth]{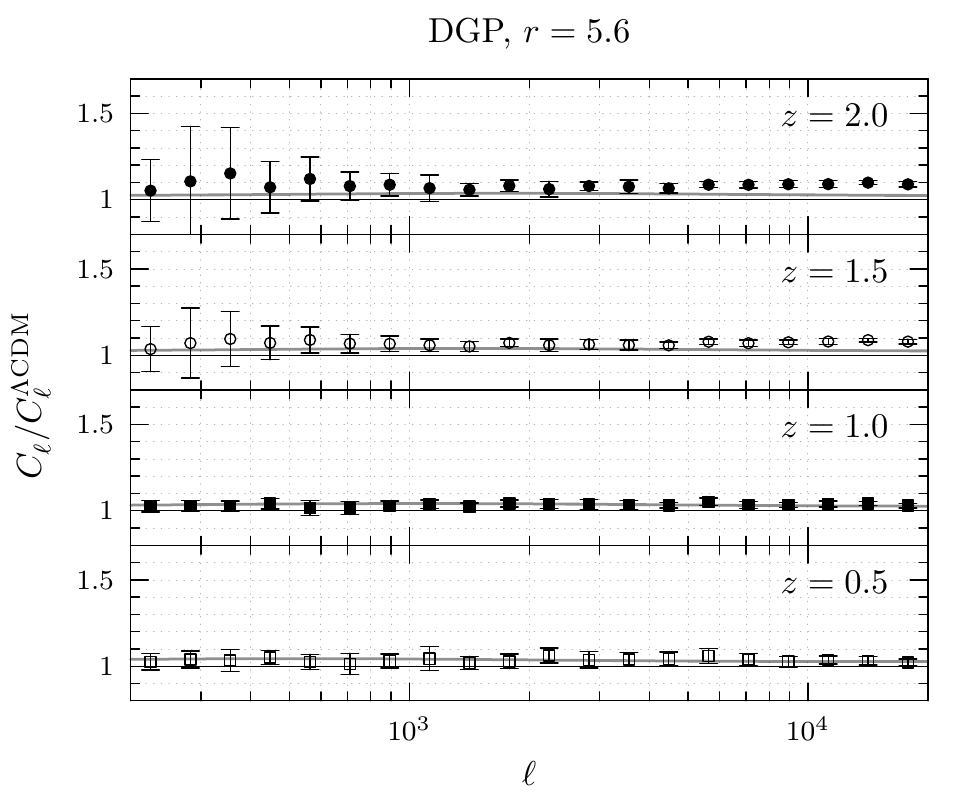}%
\caption{The fractional difference in the convergence spectra for the $f(R)$, symmetron, and DGP simulations with respect to \LCDM\ for four source redshifts.
We also show the predictions obtained from \code{halofit}.
}
\label{fig:cl_deviation}
\end{figure}

To robustly estimate $C_\ell/C_\ell^{\Lambda\mathrm{CDM}}$ at a given redshift, we calculate the convergence power spectra $C_\ell$ and $C_\ell^{\Lambda\mathrm{CDM}}$ for the same light cone taken from a screened model and the \LCDM\ simulation.
Given that all simulations start from the same initial conditions, both power spectra trace the evolution of the same patch of the universe, and the ratio $C_\ell/C_\ell^{\Lambda\mathrm{CDM}}$ is an estimator with much of the influence of cosmic variance taken out.
Only after the individual ratios $C_\ell/C_\ell^{\Lambda\mathrm{CDM}}$ have been calculated in this way do we calculate the mean and variance over the sample of light cones.
The results for the ratio $C_\ell / C_\ell^{\Lambda\mathrm{CDM}}$ can be seen in Fig.~\ref{fig:cl_deviation} for the respective $f(R)$, symmetron, and DGP models.
As intended, the sample variance has been greatly reduced, especially in the region of intermediate $\ell$.
Below we will discuss the results in more details for the three types of models individually.

\paragraph{\boldmath $f(R)$}

The convergence power spectrum in $f(R)$ shows the same qualitative behaviour as the matter power spectrum.
On large scales ($\ell \lesssim \mathcal{O}(100)$) the fractional difference w.r.t.\@ \LCDM\ is close to zero whereas for small angular scales we do find large deviations.
In the $F_6$ model the deviation from \LCDM\ is consistent (without our error bars) with zero for all source redshifts and $\ell \lesssim 10^{3}$ whereas for the $F_5$ model we have a non-zero deviation for all $\ell \gtrsim 10^2$.
The ratio $C_\ell/C_\ell^{\Lambda CDM}$ increases for both models with the angular scale and at $l\sim 10^4$ we find a $\approx 20-30\%$ ($\approx 5-10\%$) signal for the $F_5$ ($F_6$) model, depending on source redshift.

\paragraph{Symmetron}

For the symmetron simulations, we see the same trend in the convergence power spectrum as we do for $f(R)$, but with some important differences.
The fractional difference w.r.t.\@ \LCDM\ grows with $\ell$ and for $l\sim 10^4$ we find a $\approx 2 \mbox{--} 5\%$ ($\approx 20 \mbox{--} 30\%$) deviation for the $A$ ($B$) model depending on source redshift.
As opposed to the $f(R)$ model, the modifications of gravity are completely absent for $a < a_{\rm ssb}$ for the symmetron.
It is only close to the present time that the fifth-force has had time to produce a significant difference in the matter power spectrum and thus, for the simulation with $a_{\rm ssb} = 0.5$, light from high redshift sources propagates mostly through a \LCDM\ universe.
Because of this, the $C_\ell$'s for the $A$ model are very close to \LCDM\ at $z=2.0$.
As we move the source redshift closer to the present, we see more modifications to large-scale structure, and therefore larger deviations in the convergence power spectrum.

\paragraph{DGP}

In the DGP model gravity is modified even on the largest scales and we find a corresponding deviation at small values of $\ell$ down to the linear regime.
The fractional difference w.r.t.\@ \LCDM\ on large scales is found to be of similar order as in the matter power spectrum over the range $0<z<2$.
For the $r_c H_0 = 1.2$ ($r_c H_0 = 5.6$) model we find a $\approx 10 \mbox{--} 15\%$ ($\approx 5 \mbox{--} 10\%$) deviation depending on source redshift.
The deviation is, within our error bars, fairly $\ell$-independent, but we do see a slight drop-off of the signal for large $\ell$ that is consistent with the results found in the matter power spectrum (see Fig.~\ref{fig:pofk}).
Contrary to the symmetron and $f(R)$ models, we see a redshift-dependency of the signal that is close to a pure scaling of the amplitude.


\subsection{Semi-analytical predictions}

We can try and approximate the non-linear effects using semi-analytic methods such as e.g.\@ \code{halofit} that perform well in the context of \LCDM.
We find that power spectra generated through a modified version of \code{halofit} can give better agreement with the $N$-body simulations than linear theory, but even here the predictions for some models (and model parameters) can also be very poor.
This is to be expected as there is no screening included in the formalism and the \code{halofit} prediction is a function of the linear $P(k,a)$ only.
Nevertheless, for \emph{specific} models, it is possible to obtain good approximations.
There do exist other approaches for predicting the non-linear matter power spectrum (or equivalently $\Delta P / P_{\Lambda\rm CDM}$) than \code{halofit} which might do better (see for example \cite{2013PhRvD..88b3527B}), but we have chosen to focus on \code{halofit} as it is the most widely used method to obtain non-linear predictions for the matter power spectrum.
Below we will discuss the ability of \code{halofit} to predict both the matter and convergence power spectra for our three models.

\paragraph{\boldmath $f(R)$}

For $f(R)$ gravity we saw in Fig.~\ref{fig:pofk} that the linear predictions are a poor match to the simulation results for the matter power spectrum and this also holds for the convergence power spectrum.
The \code{halofit} code\footnote{\code{halofit} in this subsection refers to the modified version provided with \code{MGCAMB} \cite{2014ApJS..211...23Z}.} is able to improve significantly on this.
For the $F_5$ model, the \code{halofit} predictions for the convergence power spectrum (see Fig.~\ref{fig:cl}, \ref{fig:cl_deviation}) are roughly within the error bars for most of the $\ell$-range considered here.
Much of this good agreement can be attributed to the model-specific fitting formula used in \code{MGCAMB} to correct \code{halofit}.
This fitting formula is a $40$ parameter fit to $f(R)$ simulations, hence the good level of agreement is not surprising.
For the $F_6$ model, which is more strongly screened, we see a larger deviation for the low redshift sources and for $\ell = 10^3 \mbox{--} 10^4$ \code{halofit} predicts a $\approx 10\%$ signal whereas the simulations shows only $\approx 5\%$.

\paragraph{Symmetron}

For the symmetron there does not exist a fitting formula for the non-linear $P(k)$ to be used in \code{halofit}.
In the bottom panel of Fig.~\ref{fig:pofk} we see that \code{halofit} gives a very poor fit for the matter power spectrum.
It underpredicts the power for scales $k\lesssim 0.5~h/\Mpc$ and overpredicts the power for $k\gtrsim 0.5~h/\Mpc$.
However when we use these results to compute the convergence power spectrum, see Fig.~\ref{fig:cl},\ref{fig:cl_deviation}, we find a good agreement for the $A$ model whereas the \code{halofit} predictions show the largest deviations from the simulation results.
This is a coincidence that comes from the process of integrating the matter power spectrum to compute $C_\ell$, see Eq.~\eqref{eq:clzs}, for which the over-/underprediction of the power gets averaged out.
For the $B$ model, we see a larger deviation for the low redshift sources, just as for $f(R)$ above, and for $\ell = 10^3 \mbox{--} 10^4$ \code{halofit} predicts a $\approx 50\%$ signal whereas the simulations shows only $\approx 10\%$.

\paragraph{DGP}

For the DGP model, we find that the \code{halofit} predictions are fairly good for both the matter power spectrum and the convergence power spectrum.
The main reason for this is that the linear power spectrum is a much better (but far from good enough) approximation to the non-linear power spectrum than it is in the other two models considered here.

\medskip\noindent
In general, the \code{halofit} predictions are able to capture the signal, at least qualitatively.
However, we see a big difference for the model parameters where the screening is largest.
This illustrates how hard it is to accurately predict the signal of these models on deep non-linear scales and more elaborate methods are needed to get the accuracy needed to constrain these models using data from future weak-lensing surveys in the deep non-linear regime.


\section{Discussion}
\label{discussion}

In this paper we have investigated the weak lensing signatures of modified gravity theories that have a screening mechanism in the deep non-linear regime.
We have performed high-resolution $N$-body simulations of modified gravity models which were selected to cover the most common screening mechanisms discussed in the literature.
The output of our simulations was then processed by \code{MapSim}, which extracts randomised light cones from simulation snapshots, collapsing all particles within a given light cone onto a number of planes suitable for lensing.
These light cones contain the large-scale structure of our modified gravity theories and, using the \code{GLAMER} lensing pipeline, we created realistic convergence maps as observed by weak lensing.

Our analysis of the convergence power spectrum in screened theories of modified gravity was performed in two steps:
First, we extracted the convergence power spectrum $C_\ell$ directly from the Fourier transform of the generated convergence maps.
Our results for \LCDM\ are in good agreement with semi-analytical predictions up to $\ell \gtrsim 10^3$, as well as previous studies.
For higher values of $\ell$, we experience a loss in power which we attribute to resolution effects, while the lower values of $\ell$ have a large uncertainty due to the variance within our limited sample size.
To combat these effects, we have extracted the convergence power spectrum $C_\ell/C_\ell^{\Lambda\mathrm{CDM}}$ relative to a \LCDM\ reference simulation for each individual light cone before averaging over realisations.
Since all simulations start from the same initial conditions, this procedure reduces the influence of randomness by only considering the relative increase in power between theories of gravity.
The effects of a loss of power due to limited resolution are then mitigated and the final result is a clear indicator of the signal we can expect when searching for the deviations of screened theories from \LCDM.

For the particular modified gravity models we have simulated --- $f(R)$ gravity, the symmetron and the normal branch DGP model --- we have chosen parameters that are close to the limits that are set by local gravity experiments.
For these parameters we have found deviations of up to $\approx 50\%$ in the convergence power spectrum in the deep non-linear regime $\ell \sim 10^{4}$.
The scale ($\ell$) and source redshift dependence of the signal for these models is such that they cannot be mimicked by a \LCDM\ model with a different value of $\Omega_m/\sigma_8$ which is promising for detecting this signal in future weak lensing surveys.
However the deep non-linear regime is also where baryonic effects\footnote{%
See e.g.\@ \cite{2004APh....22..211W} for a study of baryonic effects on the convergence power spectrum.
}
and the effects of neutrinos on the power spectrum are non-negligible.
These effects can be highly degenerate with the modified gravity signal \cite{2014MNRAS.440...75B}, and a more detailed study is required to quantify these degeneracies.

We have also compared the simulation results for the matter and convergence power spectra to the predictions of linear perturbation theory and those found from applying the \code{halofit} prescription.
Performing modified gravity simulations is computationally expensive, so having an accurate prescription for predicting the modified gravity signal in the non-linear regime is of great value.
We find that linear perturbation theory (meaning the prediction for the ratio $P_{\rm MG} / P_{\Lambda\rm CDM}$ is calculated in linear theory) is a very bad approximation to the simulation results in almost all cases.
This is not surprising since linear theory does not take screening into account.
The same can be said for \code{halofit}, although it is still able to do much better than linear theory.
Nevertheless, the \code{halofit} predictions are still not good enough for precision cosmology; one can use detailed fitting formulas such as the one found in the \code{MG-halofit} code for $f(R)$ gravity, but even there the predictions can be off by as much as $\sim 5 \mbox{--} 10\%$.
Unfortunately there is no universal fitting formula with this last approach and one needs to proceed on a model by model basis; this is clearly unfeasible if one wishes to explores extensive swathes of parameter space.

This paper is a first step in laying down the tools for a thorough analysis of the effects of screening in weak lensing.
To proceed, we envision a number of steps.
For a start, we need to develop more efficient methods for generating the realisations of the density field (one approach has been advocated in \cite{2015PhRvD..91l3507W}).
Indeed multiple realisations will be needed to pin down the fine details that will allow us to distinguish between models, and an efficient scan of model parameter space is essential to be able to be place reliable constraints on screening parameters themselves. 

With such a tool in hand, we need to develop robust analytical methods (in the spirit of a modified \code{halofit}) which can be incorporated in a likelihood of up and coming data.
In particular, with the tools we have developed here, it is now possible to calibrate the weak-lensing observables of the 3-D simulations with any analytical model we choose to pursue.
An integral part of this step will be to extend our pipeline to theories where there is non-trivial gravitational slip (unlike the cases we considered here).
This will involve modifying the lensing pipeline itself to include the modified integral of the gravitational potentials.

Finally, it will be possible to focus on specific structure and go beyond the basic statistics we have looked at here.
As has been shown in \cite{2014MNRAS.442.3127S,2013PhRvL.110b1302B,2015MNRAS.451.4215Z}, voids are a promising arena; gravity will be unscreened and one expects stronger signatures of modified gravity in such a setting (see e.g. \cite{2015MNRAS.451.1036C,2014arXiv1410.8186P,2014arXiv1410.8355C,2015JCAP...08..028B,2012MNRAS.421.3481L}).


\acknowledgments

NT's and BM's research is part of the project GLENCO, funded under the European Seventh Framework Programme, Ideas, Grant Agreement n. 259349.
PGF and HAW are supported by STFC, BIPAC and the Oxford Martin School.
The calculations for this paper were performed on the DiRAC Facility jointly funded by STFC and the Large Facilities Capital Fund of BIS.
CG thanks CNES for financial support.


\appendix

\section{Field equations}
\label{field_equations}

In this appendix we present, for completeness, the field equations we solve in the $N$-body code for the different modified gravity models.
In the equations we have applied the quasi-static approximation \cite{2014PhRvD..89b3521N,2015JCAP...02..034B,2014PhRvD..89h4023L,2015PhRvD..92f4005W}.
The equation of motion for particles in our $N$-body simulations can be written in the form
\begin{equation}
\vec{\ddot x} + 2H \vec{\dot x} = -\nabla\Phi_N - \vec F_\phi,
\end{equation}
where $\nabla^2\Phi_N = 4\pi G a^2(\rho_m-\overline{\rho}_m)$ is the Newtonian potential and $\vec F_\phi$ is the fifth-force.

\paragraph{\boldmath $f(R)$}
For the Hu-Sawicky $f(R)$ model we have
\begin{equation}
\vec F_\phi = -\frac{1}{2}\nabla f_R,
\end{equation}
where the scalar field $f_R$ is determined by
\begin{equation}
\nabla^2 f_R = -\Omega_m H_0^2 a^2\left(\frac{\rho_m}{\overline{\rho}_m} - 1\right)+ a^2H_0^2\Omega_m
\left[\left(1 + 4\frac{\Omega_\Lambda}{\Omega_m}\right)\left(\frac{f_{R0}}{f_R}\right)^{n+1} - a^{-3} - 4\frac{\Omega_\Lambda}{\Omega_m}\right].
\end{equation}

\paragraph{Symmetron}
For the symmetron model we have
\begin{equation}
\vec F_\phi = \frac{6\Omega_m(H_0\lambda_\phi)^2\beta_0^2}{a_{\rm ssb}^3}\chi\nabla\chi,
\end{equation}
where the scalar field $\chi$ is determined by
\begin{equation}
\nabla^2\chi = \frac{a^2}{a\lambda_\phi^2}\left(\frac{a_{\rm ssb}^3\rho_m}{a^3\overline{\rho}_m}\chi - \chi + \chi^3\right).
\end{equation}

\paragraph{DGP}
For the DGP model we have
\begin{equation}
\vec F_\phi = \frac{1}{2}\nabla\phi,
\end{equation}
where the scalar field $\phi$ is determined by
\begin{equation}
\nabla^2\phi + \frac{r_c^2}{3\beta_{\rm DGP}(a) a^2}\left((\nabla^2\phi)^2 - (\nabla_i\nabla_j\phi)^2 \right)
= \frac{H_0^2}{a\beta_{\rm DGP}}\frac{\delta\rho_m}{\overline{\rho}_m}.
\end{equation}


\bibliographystyle{JHEP}
\bibliography{lensmg}

\end{document}